\documentclass[10pt,journal,doublecolumn]{IEEEtran}
\usepackage{cite}
\usepackage{amsmath,amssymb,placeins}
\usepackage{algorithmic}
\usepackage{graphicx}
\usepackage{textcomp}
\usepackage[mathscr]{euscript}
\usepackage{lipsum}
\usepackage{cuted}
\usepackage{mathtools}
\usepackage{multicol}
\usepackage{array}
\usepackage{makecell}
\usepackage{diagbox}
\usepackage[symbol]{footmisc}
\usepackage{caption}
\usepackage{tabularx}

\usepackage[mathscr]{euscript}
\usepackage[usenames]{color}
\usepackage{amsfonts}
\usepackage{tikz}
\usepackage{latexsym}
\usepackage{subfigure}
\usepackage[format=hang]{subfig}
\usepackage{caption}
\usepackage{floatrow}
\usepackage{multirow}
\usepackage{epstopdf}
\newtheorem{theorem}{{{\textit{Theorem}}}}

\newtheorem{lemma}{{{\textit{Lemma}}}}

\newtheorem{definition}{{{\textit{Definition}}}}

\newtheorem{remark}{{{\textit{Remark}}}}

\newtheorem{example}{{{\textit{Example}}}}

\newcommand{\Mod}[1]{\ (\mathrm{mod}\ #1)}

\def\BibTeX{{\rm B\kern-.05em{\sc i\kern-.025em b}\kern-.08em
		T\kern-.1667em\lower.7ex\hbox{E}\kern-.125emX}}
		
		\def\@fnsymbol#1{\ensuremath{\ifcase#1\or *\or \dagger\or \ddagger\or
   \mathsection\or \mathparagraph\or \|\or **\or \dagger\dagger
   \or \ddagger\ddagger \else\@ctrerr\fi}}

\begin{document}
\title{A Direct Construction of GCP and Binary CCC of Length Non-Power of Two}
\author{{Praveen Kumar, Sudhan Majhi, Subhabrata Paul}
  \thanks{Praveen Kumar and Subhabrata Paul are with the Department of Mathematics, IIT Patna, Bihta, Patna, 801106, Bihar, India (e-mail: praveen\_2021ma03@iitp.ac.in; subhabrata@iitp.ac.in). }   
  \thanks{Sudhan Majhi is with the Department of Electrical Communication Engineering, IISC Bangalore, CV Raman Rd, Bengaluru, 560012, Karnataka, India (email:smajhi@iisc.ac.in) }}

\IEEEpeerreviewmaketitle
\maketitle

%This paper presents a direct construction of Golay complementary pairs (GCPs) and binary complete complementary codes (CCCs) of non-power of two lengths. Known for their ideal correlation properties, GCPs and CCCs have found a wide range of practical applications, including coding, signal processing and wireless communication.
% GCPs and CCCs have found a wide range of practical applications including coding, signal processing and wireless communication due to their zero auto and cross-correlation sum properties. 
%In \textit{Theorem} \ref{thm1}
%\textcolor{red}
\begin{abstract}
Golay complementary pairs (GCPs) and complete complementary codes (CCCs) have found a wide range of practical applications in coding, signal processing and wireless communication due to their ideal correlation properties. In fact, binary CCCs have special advantages in spread spectrum communication due to their simple modulo-2 arithmetic operation, modulation and correlation simplicity, but they are limited in length. %while they are being adopted for LTE or 5G communication. %binary phase properties in the code. 
In this paper, we present a direct construction of GCPs, mutually orthogonal complementary sets (MOCSs) and binary CCCs of non-power of two lengths to widen their application in the recent field.
First, a generalised Boolean function (GBF) based truncation technique has been used to construct GCPs of non-power of two lengths. Then Complementary sets (CSs) and  MOCSs of lengths of the form $2^{m-1}+2^{m-3}$ ($m \geq 5$) and $2^{m-1}+2^{m-2}+2^{m-4}$ ($m \geq 6$) are generated by GBFs. Finally, binary CCCs with desired lengths are constructed using the union of MOCSs.
%A GBF based truncation is proposed to generate non-power of two length sequences. %We have also investigated the  row and column sequence peak to mean envelope power ratio (PMEPR) of the generated sequences to widen its application.
The row and column sequence peak to mean envelope power ratio (PMEPR) has been investigated and compared with existing work. The column sequence PMEPR of resultant CCCs can be effectively upper bounded by $2$. %The proposed constructions have been compared with existing work.

%Complementary sets (CS) and Golay complementary pairs (GCPs) of non power of two length are also constructed using generalised Boolean functions (GBFs) in this paper
\end{abstract}
%\IEEEpeerreviewmaketitle
\section{Introduction}\label{sec:intro}
\IEEEPARstart{T}{HE} Golay complementary pairs (GCPs) were first introduced by Golay \cite{gol1961}. The aperiodic auto-correlation sum (AACS) of a GCP diminishes to zero for all time shifts except at zero. The sequences in a GCP are known as Golay sequences. The idea of GCP is further extended to the complementary set (CS) by Tseng and Liu \cite{cliu}. A CS is a set of $M (\geq 2)$ sequences of length $N$ with the property that their AACS sum is zero for all non-zero time shifts. Tseng and Liu also proposed the concept of ($K,M,N$)-mutually orthogonal complementary set (MOCS), which is a collection of $K$ CSs %each of size $M$ and sequence length $N$
each of having  M sequences of length N, such that any two distinct CSs are orthogonal to each other, %i.e., their aperiodic cross-correlation sum (ACCS) is zero for all shifts.
%A MOCS consisting of $K$ CSs, each containing $M$ sequences of length $N$ 
%and is denoted by $(K,M,N)$-MOCS. For a MOCS, the set size $K$ is bounded above by the number of constituent sequences $M$, i.e.,
and follows the property $K\leq M$ \cite{hwa}. For a special case, when the set size of MOCS achieves its upper bound, i.e., $K=M$, it is known as a set of complete complementary code (CCC) and is denoted by ($K,K,N$)-CCC \cite{hatori}. Due to the ideal correlation properties and optimal set size, CCCs have found their application in next-generation multi-carrier code division multiple access (MC-CDMA) \cite{bell,jun,wiley,parampalli,guan}. Apart from this, CCCs are utilized in optimal channel estimation in multiple-input
and multiple-output (MIMO) frequency-selective fading channels
\cite{wang}, MIMO radar \cite{shufeng, tang}, cell search in orthogonal frequency division multiplexing (OFDM) systems \cite{min}, and data hiding\cite{kojima}.  In spread spectrum communication, the binary CCC is preferred compared to non-binary CCC due to its simple modulo-2  arithmetic operation, modulation and correlation simplicity.

Due to modulo-$2$ arithmetic operation, binary sequences are easy to implement electronically. The modulo-$2$ arithmetic is isomorphic with the use of $\{\pm 1\}$ which simplifies both the modulation and correlation processes. However, it is difficult in many cases to get flexible lengths for binary sequences. It has been proved in \cite{mjgolay} that binary GCPs exist for only even length. Binary Z-complementary pairs (ZCPs) were introduced by Fan \textit{et al.} in \cite{fan2007} and they also proved that ZCPs exists for all possible lengths. Several  constructions of binary ZCPs of different lengths are proposed in \cite{zhou2,chunlei2}. Construction of binary CSs of non-power of two lengths can be found in \cite{zhou3}.

 In the year 1999, Davis and Jedwab have proposed a direct construction of $2^h$-ary $\left(h\in \mathbb{N}\right)$ GCPs of length $2^m\left(m\in \mathbb{N}\right)$ using generalised Boolean functions (GBFs) \cite{Davis}. Paterson extended the idea of $2^h$-ary
GCPs to $q$-ary (for even $q$) GCPs \cite{pater2000}. %\textcolor{red}
%{These GBFs based construction generates GCPs having lengths of the form $2^m$ ($m \in \mathbb{N}$). So, 
%GBFs based construction of GCPs of non-power of two lengths still remains an open problem in the literature. However, using repeated application of Turyn's construction \cite{turyn}, starting with primitive GCPs of length $2,10,26$, 
The construction of GCPs of length $2^\alpha10^\beta26^\gamma$ ($\alpha,\beta,\gamma \in \mathbb{N})$ is provided by using repeated application of Turyn's construction \cite{turyn}. In \cite{pater2000}, Paterson has also proposed a GBFs based construction of CSs of length $2^m$. In the recent development GBFs based construction of CSs with more flexible lengths have been proposed in \cite{chen1,chen2,chen3,majhi,linyu}. CSs with flexible lengths are of interest to %practical orthogonal frequency division multiplexing 
OFDM systems where numbers of subcarriers are varied, i.e., non-power of two adopted by the LTE system. A direct and generalised construction of polyphase CSs is proposed in  \cite{pmz2} and it has low peak to mean envelope power ratio (PMEPR).%has applications in the OFDM system.
%On the other hand CCC with length, $2^m$ can be generated efficiently by using GBFs. 

 In \cite{arthina}, Rathinakumar and Chaturvedi proposed a direct construction of CCCs of length $2^m$ by extending the Paterson's idea of CSs generation. A number of direct constructions of CCCs with lengths $2^m$ are presented in \cite{parampalli,ccgrmc,lying2,shing2}. %All these CCCs constructions are based on GBFs and of sequence length $2^m$ only.  
 Several GBFs based constructions of Z-complementary code sets (ZCCSs) of non-power of two lengths are proposed in the literature\cite{roy,sahin,pmz,zhou4,gobinda,chunlei1}, 
to extend the number of users in ZCCS based MC-CDMA system compared to that of CCC based MC-CDMA system.
Apart from the GBFs based construction, MOCSs with non-power of two lengths can be constructed by using different systematic methods, which include reversals, negations, interleaving, concatenations etc. \cite{cliu,zhou1}. %Construction of MOCS based on the above methods can be found in . 
In the same way, Das \textit{et al.} presented  the construction of MOCSs and binary CCCs of different lengths by using paraunitary (PU) matrices \cite{das1,das2,das3}. PU matrix based construction of ZCCSs has been proposed in \cite{yubo3}. However, the sequence or code generated through these indirect methods may not be friendly for hardware generation due to their large space and time requirements.
In \cite{shing,liying} direct construction of MOCSs with non-power of two lengths have been proposed, where the set size is upper bounded by half of the number of constituent sequences in a CS, i.e., $K\leq M/2$. In addition, the authors presented an open problem of direct construction of CCCs with non-power of two lengths in \cite{shing}. Recently, Sarkar \textit{et al.} has proposed the construction of CCCs of %\textcolor{red}
{lengths $p_1^{m_1}p_2^{m_2} \cdots p_k^{m_k}$ (where $p_i$'s are prime and $m_i$'s are positive integers),} using multivariable functions (MVFs) \cite{sarkar}. This direct construction can generate $q$-ary CCCs of all possible lengths. However, in the case of $q=2$, only binary CCC of length in the form $2^m$ ($m\in \mathbb{Z}$)  can be constructed \cite{sarkar}.
So, the direct construction of GCPs and binary CCCs of non-power of two lengths is still an open problem.

By motivation of the open problem in \cite{shing,sarkar}, in this paper, direct construction of GCPs, MOCSs and binary CCCs of length $2^{m-1}+2^{m-3}$ ($m\geq5$) and $2^{m-1}+2^{m-2}+2^{m-4}$ ($m\geq 6$) have been proposed.
%In this paper, a GBFs based construction of GCPs and binary CCCs of length $2^{m-1}+2^{m-3}$ ($m\geq5$) and $2^{m-1}+2^{m-2}+2^{m-4}$ ($m\geq 6$) are proposed. This proposed construction addresses the open problems introduced in \cite{shing,sarkar}.} 
Using the idea of graphs corresponding to the quadratic part of GBFs, GCPs of non-power of two lengths are constructed. For obtaining a GCP, the graph of the quadratic part of $f$ has the property that deleting some vertices and all of their corresponding edges of the graph results in a path. In order to obtain sequences of non-power of two lengths, GBF based truncation technique is used. %We have utilized the idea of graphs corresponding to the quadratic part of the GBFs, with the property that . 
The idea has been further extended to generate CSs of non-power of two lengths. Using deleted vertices of the quadratic part of GBF, we rearrange the GBFs corresponding to the CS, and different GBF arrangements result in MOCSs of lengths non-power of two. % using the permutation of different sequences of CSs, a MOCS has been obtained.
Finally, binary CCC of non-power of two lengths has been constructed using the union of two MOCSs.
%We have also presented a direct GBFs based construction of GCPs and CSs of non power of two length using GBFs.

The remaining paper is organized as follows. Basic notations and definitions are provided in Section II. In sections III, IV and V the constructions of GCPs, MOCSs and binary CCCs of length non-power of two are given respectively. Section VI describes how to build additional non-power two-length GCPs, MOCSs, and CCCs. Section VII provides the row and column sequence PMEPR of the proposed CCCs. Finally, concluding remarks are provided in Section VIII.
%\vspace{-1cm}
\section{Notations and Definitions}\label{sec2}
%\vspace{-0.5cm}
%\textcolor{red}
{In this section, the preliminaries, notations, and immediate results required for our proposed construction are discussed.}
\begin{definition}\label{def1} Let $\mathbf{d}$ $=(d_{0},d_{1}, \ldots, d_{N-1})$ and $\mathbf{e}$ $=(e_0,e_1, \ldots, e_{N-1})$ be two length $N$  complex-valued sequences then the aperiodic cross-correlation function 
(ACCF) between $\mathbf{d}$ and $\mathbf{e}$ at a shift $s$ ($s \in \mathbb{Z}$) can be defined as
\begin{equation}
\mathcal{C}\left(\mathbf{d}, \mathbf{e}\right)(s)=\begin{cases}
\sum_{k=0}^{N-1-s} d_{k}\cdot e^*_{k+s}, & 0 \leq s \leq N-1, \\
\sum_{k=0}^{N-1-s} d_{k+s}\cdot e^*_{k}, & -N+1 \leq s \leq-1,\\
0, & |s| \geq N,
\end{cases}
\end{equation}
%\end{definition}
where $()^*$ is the complex conjugate operator. When $\mathbf{d}$ and $\mathbf{e}$ are equal, it is known as aperiodic auto-correlation function 
(AACF) of $\mathbf{e}$ and is denoted by $\mathcal{A}(\mathbf{\mathbf{e})}(s).$\\
We can also define the ACCF and AACF of $\mathbb{Z}_q$ valued sequences by defining a one-one correspondence between $\mathbb{Z}_q$ valued sequence $\mathbf{e}$=$(e_0,e_1, \ldots, e_{N-1})$ and the complex-valued sequence $\mathbf{e'}$=$(e'_0,e'_1, \ldots, e'_{N-1})$, where $e'_i=\omega^{e_i}$ and $\omega=\exp\left(2\pi \sqrt{-1}/q\right)$ is $q$th root of unity. So if $\mathbf{d}$ and $\mathbf{e}$ are $\mathbb{Z}_q$ valued sequences then we define their ACCF $\mathcal{C}(\mathbf{d,e})(s)$ and AACF  $\mathcal{A}(\mathbf{e})(s)$ respectively as ACCF and AACF of the corresponding complex-value sequence $\mathbf{d'}$ and $\mathbf{e'}$.
\end{definition}
\begin{definition}\label{def2} A set of $M$ sequences $\mathbf{e}^{0}, \mathbf{e}^{1}, \ldots, \mathbf{e}^{M-1}$, each of length $N$, is said to be a CS if
$$
\begin{gathered}
\mathcal{A}\left(\mathbf{e}^{0}\right)(s)+\mathcal{A}\left(\mathbf{e}^{1}\right)(s)+\cdots+\mathcal{A}\left(\mathbf{e}^{M-1}\right)(s) \\
= \begin{cases}MN, & s=0, \\
0, & \text { otherwise .}\end{cases}
\end{gathered}
$$
For $M=2$, it is known as a
%Golay complementary pair
GCP.
\end{definition}
\begin{definition}\label{def3} 
 Consider a set  $\mathcal{E}=\left\{{E^{0}}, {E^{1}}, \cdots, {E^{K-1}}\right\}$, where each set ${E^{p}}$ consists of $M$ sequences, i.e., ${E^{p}}=\left\{\mathbf{e}_{0}^{p}, \mathbf{e}_{1}^{p}, \cdots, \mathbf{e}_{M-1}^{p}\right\}$, and length of each sequence $\mathbf{e}_{l}^{p}$ is $N$, where $0 \leq p \leq K-1$ and $0 \leq l \leq M-1$. The set $\mathcal{E}$ is called an MOCS, denoted by $(K,M,N)$-MOCS, if the ACCF of $E^{p}$ and $E^{p^{\prime}}$ satisfies
\begin{equation}\label{eqn1}
\begin{aligned}
\mathcal{C}\left(E^{p}, E^{p^{\prime}}\right)(s)&=\sum_{n=0}^{M-1} \mathcal{C}\left(\mathbf{e}_{n}^{p}, \mathbf{e}_{n}^{p^{\prime}}\right)(s)\\&
= \begin{cases}MN, & s=0, p=p^{\prime}, \\ 0, & \text { otherwise },\end{cases}
\end{aligned}
\end{equation}
where $0 \leq p, p^{\prime} \leq K-1;$ $K,M$ and $N$ are known as the set size, flock size and sequence length respectively.
 For a $(K, M, N)$-MOCS, the set size is always smaller than the flock size, i.e., $K \leq M$. For the special case when $K=M$, the MOCS is called a CCC of order $K$ and length $N$, and is denoted by $(K,K,N)$-CCC.
 \end{definition}
\subsection{Generalised Boolean function}
A GBF $f$ in $m$ binary variables $y_0,y_1,\hdots,y_{m-1}$ is a function from $\{0,1\}^m$ to $\mathbb{Z}_q$, where $q\geq 2$ is an even integer. A monomial of degree $r$ is defined as the product of any $r$ variables among $y_0,y_1,\hdots,y_{m-1}$. So there are $\sum_{r=0}^{m}{m\choose r}$ = $2^m$ monomials, namely $1,y_0,y_1$, $\hdots,y_{m-1},y_0y_1,y_0y_2,\hdots,y_{m-2}y_{m-1},\hdots,y_0y_1\cdots y_{m-1}$. With the linear combination of these $2^m$ monomials and by taking coefficient from $\mathbb{Z}_q$, a GBF can be expressed uniquely. In the expression of a GBF of order $r$, there exist at least one highest-degree monomial of order $r$ with non-zero coefficient. Corresponding to a GBF $f$ of $m$ variables $y_0,y_1,\hdots,y_{m-1}$, length $2^m$ $\mathbb{Z}_q$-valued vector is expressed as 
\begin{equation}\label{eqn2}
 \mathbf{f}=\left({f_0}, {f_1}, \hdots, {f_{2^m-1}}\right),
\end{equation}
where $f_i=f(i_0,i_1,\hdots,i_{m-1})$ 
and $(i_0,i_1,\hdots,i_{m-1})$ is the 
binary vector representation of $i$.
%where as in the remainder of this letter, $q$ is an even integer not less than $2$. 
A complex-valued vector $\mathbf{f'}$ is associated with every $\mathbf{f}$ by ${f}'_i=\omega^{f_i}$. %and $\omega=\exp\left(2\pi\sqrt{-1}/q\right)$.
When it is clear from the context, only $f$ is used to refer to both.
%we will just use $f$ to refer to all three.
Corresponding to a GBF $f$ with $m$ variables the sequence $\mathbf{f}$ is of length $2^{m} .$\\

We can restrict the domain of GBF to get sequences of length non-power of two. %We define a GBF in variables $m$ binary variables $y_0,y_1,\hdots,y_{m-1}$ as $f:A \rightarrow \mathbb{Z}_q$, 
Let us define a set $A$ which is a subset of $\{0,1\}^m$. So, depending upon the domain $A$ we can get different length sequences corresponding to GBF $f$. By $\mathbf{a}_m$ we mean the binary vector representation of positive integer $\mathbf{a}$ in $m$ components.
\begin{example}\label{example0}
Let $f:A \rightarrow \mathbb{Z}_2$ be defined as $f(y_0,y_1,y_2)=y_0y_1+y_2$, where $A=\{\mathbf{0}_3,\mathbf{1}_3,\hdots,\mathbf{5}_3\}$, then the sequence corresponding to $f$ is  $\left(1,1,1,-1,-1,-1\right)$, which is of length $6$.
Similarly, if we define $A= \{\mathbf{3}_3,\mathbf{4}_3,\hdots,\mathbf{7}_3\}$, then we get the sequence $\left(-1,-1,-1,-1,1\right)$, which is of length $5$.
\end{example}
%In this paper, we focus on sequences of length $\neq 2^{m} .$ Hence, we define the truncated sequence  $\mathbf{f}^{(L)}$ and $\mathbf{f}_{(L)}$  corresponding to the GBF $f$ by  deleting the first $L$ and last $L$ elements of the sequence $\mathbf{f}$ respectively. For the sake of simplicity, we ignore the superscript of $\mathbf{f}^{(L)}$ and subscript of $\mathbf{f}_{(L)}$ when the sequence length is known from the context. Also for a set $A$ containing  sequences of equal length, $\Psi^L(A)$ and $\Psi_L(A)$ represents a set by deleting the first $L$ and last $L$ elements of each sequence of the set $A$ respectively. 

{
\subsection{Graph of Quadratic form of GBF}
%Consider a set 
%$$\mathcal{P}=\left\{\sum \prod_{0 \leq i \leq m-1} x_i^{p_i}:p_i\in \{0,1\}, 3\leq \sum p_i \leq m\right\}.$$
Let $Q:\{0,1\}^{m} \rightarrow \mathbb{Z}_{q}$ be a GBF of order $2$ defined by
\begin{equation}
Q\left(y_{0},y_1, \hdots, y_{m-1}\right)= \sum_{0 \leq i<j<k} q_{i j} y_{i} y_{j},
\end{equation}
%where $\mathcal{H}$ represents the linear combination of elements of set $\mathcal{P}$, with coefficients from $\mathbb{Z}_{q}$ and 
%$Q$ is the quadratic form of the GBF, given by
 where $k\leq m$ and $q_{i j} \in \mathbb{Z}_{q}$}. We associate a labeled graph $G(Q)$ corresponding to the GBF $Q$, on $k$ vertices %For defining the correspondence, we only use the quadratic part $Q$ of the GBF $f$. So by $G(f)$ we actually mean $G(Q)$.
 by representing the vertices of $G(Q)$ by $0,1, \hdots, m-1$ and joining two vertices $i$ and $j$ by an edge labeled $q_{i j}$ if and only if $q_{i j} \neq 0$. In the case, $q=2$, $q_{ij}$ can only take values either $0$ or $1$, so every edge is labelled 1 and by convention, edge labels are omitted in this case . From any given graph $G(Q)$ of this type, the quadratic form $Q$ can be easily and uniquely recovered. % If $f:\{0,1\}^{m} \rightarrow$ $\mathbb{Z}_{q}$ is a GBF of order $r$, then we define $G(f)$ to be the graph $G(Q)$ where $Q$ is the quadratic part of $f$. 
A graph $G(Q)$ is called a path on $k$ vertices if the number of edges is exactly one less than the number of vertices, and each edge is labelled $q/2$. For $k=1$, this is a trivial path and for $k\geq 2$, this type of path is known as the Hamiltonian path.
%\begin{itemize}
   % \item $m=1$ (in which case the graph contains a single vertex and no edges), or
%\item $m \geq 2$ and $G$ has exactly $m-1$ edges, all labeled $q / 2$ which form a Hamiltonian path in $G$.
%\end{itemize}
  For $2\leq k < m$, a path on $k$ vertices corresponds to a quadratic form of the type
\begin{equation}
\frac{q}{2} \cdot \sum_{\alpha=1}^{k-1} y_{\pi(\alpha-1)} y_{\pi(\alpha)},
\end{equation}
where $\pi$ is a permutation of the set $\{0,1, \hdots, k-1\}$.

\subsection{Restricted Boolean function}

Let $f:A\subseteq\{0,1\}^{m} \rightarrow \mathbb{Z}_{q}$ be a GBF in variables $y_{0}, y_{1}, \hdots, y_{m-1}$ and $\mathbf{y}=\left(y_{p_{0}} y_{p_{1}} \cdots y_{p_{k-1}}\right)$ where $0 \leq p_{0}<p_{1}<\cdots<p_{k-1}<m$ . Let $\mathbf{c}=$ $\left(c_{0} c_{1} \cdots c_{k-1}\right)$ be a binary word of length $k$, i.e., $c_i\in \{0,1\}$. Then the vector $f|_{\mathbf{y=c}}$ is defined to be the complex-valued vector with component $i=\sum_{j=0}^{m-1} i_{j} 2^{j}$ equal to $\omega^{f\left(i_{0}, i_{1}, \hdots, i_{m-1}\right)}$ if $i_{j_{\alpha}}=c_{\alpha}$ for each
$0 \leq \alpha<k$, and equal to 0 otherwise. %Here $\omega$ is a complex $q$th root of unity.
%By convention, in the special case where $\mathbf{x}$ and $\mathbf{c}$ are null (i.e., of length zero), we define $f|_{x=c}$ to be the complex-valued vector associated with $f$.
%The complex-valued vector associated with $f$ is defined by convention as $f|_{x=c}$ when $\mathbf{x}$ and $\mathbf{c}$ are null (i.e., of length 0).
As a convention, if $\mathbf{y}$ and $\mathbf{c}$ are null (i.e., of length 0), then $f|_{\mathbf{y=c}}$ represents the complex-valued vector associated with $f$.
%A simple consequence of the definition is the following vector identity: for any $\mathbf{x}$ defined as above
%$$f=\sum_{c} f|_{x=c}.$$ 
%The vector $f|_{x=c}$ can also be thought of as being constructed from the GBF $f$ by substituting $x_{i_{\alpha}}=$ $c_{\alpha}$ in the algebraic normal form for $f$ for each $0 \leq \alpha<k$, simplifying to obtain a GBF in $m-k$ variables (which we also denote by $\left.\left.f\right|_{x=c}\right)$, then calculating $\omega^{\left.f\right|_{x=c}}$ over the domain of this new function and finally inserting zeros at appropriate locations in the resulting vector.
\begin{lemma}[\cite{arthina}]\label{lem1}
 Let $f, g:A\subseteq\{0,1\}^{m} \rightarrow \mathbb{Z}_{q}$ be GBFs in variables $y_{0}, y_{1}, \hdots, y_{m-1}$. Let $\mathbf{y}=\left(y_{p_{0}} y_{p_{1}} \cdots y_{p_{k-1}}\right)$ where $0 \leq p_{0}<p_{1}<\cdots<p_{k-1}<m$ and $\mathbf{c}=$ $\left(c_{0} c_{1} \cdots c_{k-1}\right)$ be a binary word of length $k$. Further let us denote $\mathbf{z}=\left(z_{i_{0}} z_{i_{1}} \cdots z_{i_{l-1}}\right)$ where  $0 \leq i_{1}<i_{2}<\cdots<$ $i_{l-1}<m$ be a set of indices not in $\left\{p_{0}, p_{1}, \ldots, p_{k-1}\right\}$. Then for a binary vector $\mathbf{n}=$ $\left(n_{0} n_{1} \cdots n_{k-1}\right)$, the following equality holds
\begin{equation}
\mathcal{C}\left(\left.f\right|_{\mathbf{y}=\mathbf{c}},\left.g\right|_{\mathbf{y}=\mathbf{n}}\right)(s)=\sum_{\mathbf{c}_{1}, \mathbf{c}_{2}} \mathcal{C}\left(\left.f\right|_{\mathbf{y z}=\mathbf{c c}_{1}},\left.g\right|_{\mathbf{y z}=\mathbf{n} \mathbf{c}_{2}}\right)(s).
\end{equation}
\end{lemma}
\begin{lemma}[\cite{pater2000}]\label{lem2}
Let $f:A\subseteq\{0,1\}^{m} \rightarrow \mathbb{Z}_{q}$ be a GBF in variables $y_{0}, y_{1}, \cdots, y_{m-1}$. Let $\mathbf{y}$ and $\mathbf{c}$ are as defined in \textit{Lemma} \ref{lem1}, then AACF is given by
\begin{equation}
\mathcal{A}(f)(s)=\sum_{\mathbf{c}} \mathcal{A}\left(\left.f\right|_{\mathbf{y=c}}\right)(s)+\sum_\mathbf{{c}_{1} \neq \mathbf{c}_{2}} \mathcal{C}\left(\left.f\right|_\mathbf{{y=c}_{1}},\left.f\right|_\mathbf{{y=c}_{2}}\right)(s).
\end{equation}
\end{lemma}
%\textcolor{red}{
%\subsection{Peak to mean envelope power ratio (PMEPR)}
%In an OFDM system, sequences with low PMEPR are desirable due to their low power consumption in transmission. A short introduction of PMEPR is discussed in this subsection.\\
%Consider an OFDM signal with $n$ sub-carriers, with carrier frequency and sub-carrier spacing as $f_0$ and $\Delta f$, respectively. The OFDM signal of a sequence  $\mathbf{e}$ $=[e_0,e_1, \ldots, e_{N-1}]$ is defined as the real part of 
$$
%S_{\mathbf{e}}(t)=\sum_{k=0}^{N-1} e_{k} e^{2 \pi j\left(f_{0}+k \Delta f\right) t},~~ 0 \leq {\Delta f}t<{1},
$$
%where $j$ is the square root of unity. Then the PMEPR of $\mathbf{e}$ is defined as,
$$
%P M E P R(\mathbf{e})=\frac{1}{N} \sup _{0 \leq {\Delta f}t<{1}}\left|S_{\mathbf{e}}(t)\right|^{2}.
$$
%The PMEPR of a GCP and CS of set size $K$ is bounded above by $2$ and $K$ respectively \cite{pater2000,Davis}.}
\section{Proposed Construction of GCPs}\label{sec3}
In this section, we provide a GBFs based construction of GCPs for non-power of two lengths. 
Unless otherwise stated, this section and subsequent sections assume $m \geq 5$.

Suppose $Q:\{0,1\}^{m-4} \rightarrow \mathbb{Z}_q$ is the quadratic form in variables $z_0,z_1,\hdots,z_{m-5}$, i.e.,
\begin{equation}\label{eqn5}
    Q\left(z_0,z_1,\hdots,z_{m-5}\right)=\sum_{0\leq i < j < m-4} q_{ij}z_iz_j.
\end{equation}
For any $c,c_i \in \mathbb{Z}_q$, we define a GBF
\begin{equation}\label{eqn6}
    f_1=Q+\sum_{i=0}^{m-5}c_iz_i+c.
\end{equation}
%where $Q$ is defined in (\ref{eqn5}).
Using the notation $\bar z_i=1-z_i$ and $f_1$ defined in (\ref{eqn6}), the proposed GBF $f:A \rightarrow \mathbb{Z}_q$ is defined as 
\begin{equation}\label{eqn7}
\begin{aligned}
    f=&f_1+\frac{q}{2}\bar z_{m-1}\left(\bar z_{m-4}\left(z_{m-3}+z_{m-2}\right)+z_{m-2}z_{m-3}\right)\\&
    +\frac{q}{2}z_{\beta_1}\left(\bar z_{m-1}\left(z_{m-2}\bar z_{m-3}\bar z_{m-4}+z_{m-2}z_{m-3}\right)\right.\\& \left.+z_{m-1}\bar z_{m-2}\bar z_{m-3} \right),
\end{aligned}
\end{equation}
where $A=\left\{\mathbf{0}_m,\mathbf{1}_m,\hdots(\mathbf{2^{m-1}+2^{m-3}-1})_m\right\}$. 

We will first prove a special case when the quadratic part of $f$ given in (\ref{eqn7}) is zero.
\begin{lemma}\label{lem30}
Let the quadratic part $Q$ of $f|_\mathbf{z=c}$ be identically equal to zero and $G\left(\left.Q\right|_\mathbf{z=c}\right)$ has a single vertex labeled $\beta$, where $\mathbf{z}=(z_{p_0},z_{p_1},\hdots,z_{p_{m-6}})$ and $\mathbf{c}=(c_0c_1\cdots c_{m-6})$ be a $(m-5)$ length binary vector. Then
\begin{equation}
\left(f|_\mathbf{z=c},\left(f+\frac{q}{2}z_{\beta}+c'\right)|_\mathbf{z=c}\right),
\end{equation}
forms a GCP of length $2^{m-1}+2^{m-3}$, with exactly $20$ non-zero elements.
\end{lemma}
\begin{IEEEproof}
%For $k=m-5$, in this case the quadratic part $Q$, of $f|_\mathbf{x=c}$ is identically zero and $G\left(\left.f\right|_\mathbf{x=c}\right)$ has a single vertex labeled $\beta$, and $\mathbf{x}$ omits exactly the variable $x_\beta$ among \{$x_0,x_1,\hdots,x_{m-5}$\}.
Since $f_1|_\mathbf{z=c}$ is a function containing only one variable $z_\beta$, so $f_1|_\mathbf{z=c}$ gives exactly $2$ non-zero elements in the sequence. The binary variables $z_{m-4},z_{m-3},z_{m-2}$ and $z_{m-1}$ remain unaffected by $\mathbf{z=c}$, and since the length of the sequence is $10 \times 2^{m-4}$, so the function $f|_\mathbf{z=c}$ takes non-zero values in exactly 20 components numbered %The higher order terms of the function $f|_\mathbf{z=c}$ represent concatenation of $10$ sequences each having exactly $2$ non-zero terms,  where the $1$st and $2$nd non-zero component of the $(k+1)$th sequence occurs at position numbered
%So the function $f|_\mathbf{z=c}$ takes non-zero values in exactly 20 components numbered 
$ k2^{m-4}+\sum_{j\neq \beta}c_j2^j$, $0\leq k \leq 9$ and  $2^\beta+k2^{m-4}+\sum_{j\neq \beta}c_j2^j $, $0\leq k \leq 9$.
 These non-zero terms are placed in increasing order as follows
\begin{equation*}
\begin{aligned}
&\left\{\omega^{\gamma},\omega^{\delta},\omega^{\gamma},\omega^{\delta},-\omega^{\gamma},-\omega^{\delta},\omega^{\gamma},\omega^{\delta},-\omega^{\gamma},\omega^{\delta}\right. \\&  \left. ,\omega^{\gamma},\omega^{\delta},-\omega^{\gamma} ,\omega^{\delta},-\omega^{\gamma},\omega^{\delta},\omega^{\gamma},-\omega^{\delta},\omega^{\gamma},-\omega^{\delta}\right\},
\end{aligned}
\end{equation*}
where $\gamma$ and $\delta$ are the values taken by the function $f_1|_\mathbf{z=c}$ at $\sum_{j\neq \beta}c_j2^j$ and  $2^\beta+\sum_{j\neq \beta}c_j2^j$ respectively.

The function $\left(f_1+\frac{q}{2}z_\beta+c'\right)|_\mathbf{z=c}$
 takes values $\gamma+c'$ and $\delta+\frac{q}{2}+c'$ at positions $\sum_{j\neq \beta}c_j2^j$ and  $2^\beta+\sum_{j\neq \beta}c_j2^j$ respectively. So the 20 non-zero components of the function $\left(f+\frac{q}{2}z_{\beta}+c'\right)|_\mathbf{z=c}$ at positions mentioned above are placed in increasing order as follows
\begin{equation*}
\begin{aligned}
&\left\{\omega^{\gamma+c'},-\omega^{\delta+c'},\omega^{\gamma+c'},-\omega^{\delta+c'},-\omega^{\gamma+c'},\omega^{\delta+c'},\omega^{\gamma+c'}\right. \\& \left.,-\omega^{\delta+c'},-\omega^{\gamma+c'},-\omega^{\delta+c'} ,\omega^{\gamma+c'},-\omega^{\delta+c'},-\omega^{\gamma+c'}\right.\\& \left.,-\omega^{\delta+c'},-\omega^{\gamma+c'},-\omega^{\delta+c'},\omega^{\gamma+c'},\omega^{\delta+c'},\omega^{\gamma+c'},\omega^{\delta+c'}\right\}.
\end{aligned}
\end{equation*}
%Now it can be easily verified through direct calculation that $f|_\mathbf{x=c}$ and $\left(f+x_{\beta_2}+c'\right)|_\mathbf{x=c}$ forms a GCP.\\
The non-zero value of the AACF of the vectors corresponding to $f|_\mathbf{z=c}$ and $\left(f+\frac{q}{2}z_{\beta}+c'\right)|_\mathbf{z=c}$ occurs only at shifts $s=k2^{m-4}+2^\beta$, $0\leq k \leq 9$ and $s=k2^{m-4}-2^\beta$, $1\leq k \leq 9$.\\
For $s=k2^{m-4}+2^\beta$ and $0\leq k \leq 9$, the AACF of the above two functions are expressed as 
\begin{equation}
\mathcal{A}\left(f|_\mathbf{z=c}\right)\left(s\right)= \mathrm{t}_k\omega^{\gamma}\left(\omega^{\delta}\right)^*= \mathrm{t}_k\omega^{\gamma-\delta},
\end{equation}
and 
\begin{equation}
\begin{aligned}
\mathcal{A}\left(\left(f+\frac{q}{2}z_{\beta}+c'\right)|_\mathbf{z=c}\right)\left(s\right)&= \mathrm{t}_k\omega^{\gamma+c'}\left(-\omega^{\delta+c'}\right)^*\\&=-\mathrm{t}_k\omega^{\gamma-\delta},
\end{aligned}
\end{equation}
where $\mathrm{t}_k$ is some constant.

Similarly for $s=k2^{m-4}-2^\beta$ and $1\leq k \leq 9$, the above can be written as 
\begin{equation}
\mathcal{A}\left(f|_\mathbf{z=c}\right)\left(s\right)= \mathrm{t'}_k\omega^{\delta}\left(\omega^{\gamma}\right)^*= \mathrm{t'}_k\omega^{\delta-\gamma},
\end{equation}
and 
\begin{equation}
\begin{aligned}
\mathcal{A}\left(\left(f+\frac{q}{2}z_{\beta}+c'\right)|_\mathbf{z=c}\right)\left(s\right)&= \mathrm{t'}_k(-\omega^{\delta+c'})\left(-\omega^{\gamma+c'}\right)^*\\&=-\mathrm{t'}_k\omega^{\delta-\gamma},
\end{aligned}
\end{equation}
where $\mathrm{t'}_k$ is some constant. So the AACS is zero for all $s \neq 0$, and hence the result follows.
\end{IEEEproof}
Some notations are defined below for proving the general case of construction of GCPs of non-power of two length. Let $0\leq p_0<p_1<\cdots<p_{k-1}<m-4$, be a list of $k$ indices, where $0\leq k \leq m-5$ and $\mathbf{z}=(z_{p_0},z_{p_1},\hdots,z_{p_{k-1}})$.
Let the remaining $m-4-k$ indices between 0 to $m-5$  be $0\leq i_0<i_1<\cdots<i_{m-k-5}<m-4$.
Let $\mathbf{c}=(c_0c_1\cdots c_{k-1})$ be a $k$ length binary vector.
\begin{theorem}\label{thm1}
Let us consider the restricted function $f|_\mathbf{z=c}$ that is obtained by restricting the variables $z_{p_\alpha}, 0\leq \alpha \leq k \leq m-5$, of GBF $f$ in (\ref{eqn7}) with the property that
$G\left(\left.Q\right|_\mathbf{z=c}\right)$ is a path. Let  $\beta_1$ and $\beta_2$ be the two end vertices of the path
$G\left(\left.Q\right|_\mathbf{z=c}\right)$ when $0\leq k < m-5$. In case of $k=m-5$, $G\left(\left.Q\right|_\mathbf{z=c}\right)$ has only a single vertex labeled $\beta=\beta_1=\beta_2$. Then for any $c'\in \mathbb{Z}_q$, the complex-valued vectors $f|_\mathbf{z=c}$ and $\left(f+\frac{q}{2}z_{\beta_2}+c'\right)|_\mathbf{z=c}$ forms a GCP of length $2^{m-1}+2^{m-3}$.
\end{theorem}
\begin{IEEEproof}
We prove the result using induction on $k$, where the statement of the theorem is taken as an inductive hypothesis.
The case when $k=m-5$, follows directly from \textit{Lemma} \ref{lem30}.
Now, let the theorem be true when $\mathbf{z}$ contains $k+1$ variables, and we consider the case for $k$ variables, where $0\leq k< m-5$. % When $G\left(\left.Q\right|_\mathbf{z=c}\right)$ is a path, values of function $f|_\mathbf{z=c}$ in variables $\left(z_{i_0},z_{i_1},\hdots,z_{i_{m-k-5}},z_{m-4},z_{m-3},z_{m-2},z_{m-1}\right)$ determine the non-zero components of function $f$.
When $G\left(\left.Q\right|_\mathbf{z=c}\right)$ is a path, the non-zero components of function $f$ are determined by the values of function $f|_\mathbf{z=c}$ in variables $\left(z_{i_0},z_{i_1},\hdots,z_{i_{m-k-5}},z_{m-4},z_{m-3},z_{m-2},z_{m-1}\right)$.

%The non-zero components of the function $f|_\mathbf{x=c}$ are determined by the values of the function $f|_\mathbf{x=c}$ in variables $\left(x_{i_0},x_{i_1},\hdots,x_{i_{m-k-5}},x_{m-4},x_{m-3},x_{m-2},x_{m-1}\right)$, . 
So for some permutation $\pi$ of $\{0,1,\hdots,m-k-5\}$ and $c_0,c_1,\hdots,c_{m-k-5},c \in \mathbb{Z}_q$, we get the function
\begin{equation}\label{eqn9}
\begin{aligned}
    &f|_\mathbf{z=c}\left(z_{i_0},z_{i_1},\hdots,z_{i_{m-k-5}},z_{m-4},z_{m-3},z_{m-2},z_{m-1}\right)\\=& \frac{q}{2}\sum_{\alpha=0}^{m-k-6}z_{i_{\pi(\alpha)}}z_{i_{\pi(\alpha+1)}}+\sum_{\alpha=0}^{m-k-5}c_{\alpha}z_{i_{\pi(\alpha)}}+c\\&
    \!+\!\frac{q}{2}z_{i_{\pi(m-k-5)}}\!\left(\bar z_{m-1}\!\left(z_{m-2}\bar z_{m-3}\bar z_{m-4}\!+\!z_{m-2}z_{m-3}\right)\!+\!z_{m-1} \right. \\& \left.\bar z_{m-2}\bar z_{m-3} \right)\!+\! \frac{q}{2}\bar z_{m-1}\left(\bar z_{m-4}\left(z_{m-3}\!+\!z_{m-2}\right)\!+\!z_{m-2}z_{m-3}\right).
    \end{aligned}
\end{equation}
The higher order terms in (\ref{eqn9}) is utilized frequently, so for simplicity, it is denoted by $R$ as follows
\begin{equation}\label{eqn10}
    \begin{aligned}
        R\!=&\frac{q}{2}z_{i_{\pi(m-k-5)}}\!\left(\bar z_{m-\!1}\!\left(z_{m-\!2}\bar z_{m-\!3}\bar z_{m-\!4}\!+\!z_{m-\!2}z_{m-\!3}\right)\!+\!z_{m-\!1}\! \right. \\& \left.\!\bar z_{m-2}\bar z_{m-3} \right)\!+\! \frac{q}{2}\bar z_{m-1}\left(\bar z_{m-4}\left(z_{m-3}\!+\!z_{m-2}\right)\!+\!z_{m-2}z_{m-3}\right).
    \end{aligned}
\end{equation}
Now, the aim is to prove that the sequences $f|_\mathbf{z=c}$ and $\left(f+\frac{q}{2}z_{i_{\pi(0)}}+c'\right)|_\mathbf{z=c}$, where $c'\in \mathbb{Z}_q$ is arbitrary, forms a GCP of length $2^{m-1}+2^{m-3}$.
If $s \neq 0 $ is chosen arbitrarily, then the sum of AACF of the sequences is given by
\begin{equation}\label{eqn11}
\begin{aligned}
   & \mathcal{A}\left(f|_\mathbf{z=c}\right)\left(s\right)+  \mathcal{A}\left(\left(f+\frac{q}{2}z_{i_{\pi(0)}}+c\right)|_\mathbf{z=c}\right)\left(s\right)\\=&\mathcal{A}(g_1)(s)+\mathcal{A}(g_2)(s)+\mathcal{C}(g_1,g_2)(s)+\mathcal{C}(g_2,g_1)(s)\\&+\mathcal{A}(g_3)(s)+\mathcal{A}(g_4)(s)+\mathcal{C}(g_3,g_4)(s)+\mathcal{C}(g_4,g_3)(s),
    \end{aligned}
\end{equation}
where $g_1\!=\!f|_{\mathbf{z}z_{i_{\pi(0)}}=\mathbf{c}0}$, 
$g_3=\left(f+\frac{q}{2}z_{i_{\pi(0)}}+c'\right)|_{\mathbf{z}z_{i_{\pi(0)}}=\mathbf{c}0}$,\\
$~~~~~~~~~~g_2=f|_{\mathbf{z}z_{i_{\pi(0)}}=\mathbf{c}1}$,
$g_4=\left(f+\frac{q}{2}z_{i_{\pi(0)}}+c'\right)|_{\mathbf{z}z_{i_{\pi(0)}}=\mathbf{c}1}$.\\
The non-zero components of the vector $g_1$ are derived from a function $h_1$ by substituting $z_{i_{\pi(0)}}=0$ in the function $f|_\mathbf{z=c}$ in (\ref{eqn9}). For $0 \leq k \leq m-7$, the function $h_1$ is given by
\begin{equation}
    \begin{aligned}
       & h_1|_\mathbf{z=c}\left(\!z_{i_{\pi(0)}},\!z_{i_{\pi(1)}},\!\hdots,\!z_{i_{\pi(m-k-5)}},\!z_{m-4},\!z_{m-3},\!z_{m-2},\!z_{m-1}\!\right)\\&= \frac{q}{2}\sum_{\alpha=1}^{m-k-6}z_{i_{\pi(\alpha)}}z_{i_{\pi(\alpha+1)}}+\sum_{\alpha=1}^{m-k-5}c_{\alpha}z_{i_{\pi(\alpha)}}+c+R.
    \end{aligned}
\end{equation}
While for $k=m-6$, it is given by
\begin{equation}
    \begin{aligned}
       &h_1\left(z_{i_{\pi(0)}},z_{i_{\pi(1)}},\hdots,z_{i_{\pi(m-k-5)}},z_{m-4},z_{m-3},z_{m-2},z_{m-1}\right)\\&=c_1z_{i_{\pi(1)}}+c+R,
    \end{aligned}
\end{equation}
%where $R$ is defined in (\ref{eqn10}).\\
Similarly, by substituting $z_{i_{\pi(0)}}=1$ in the function $f|_\mathbf{z=c}$, function $h_2$ is obtained which yields the non-zero components of the vector $g_2$. The function $h_2$ is given by 
\begin{equation}
    \begin{aligned}
       &h_2\left(z_{i_{\pi(0)}},z_{i_{\pi(1)}},\hdots,z_{i_{\pi(m-k-5)}},z_{m-4},z_{m-3},z_{m-2},z_{m-1}\right)\\&= h_1+z_{i_{\pi(1)}}+c_0.
    \end{aligned}
\end{equation}
To easily calculate the AACF of $g_2$, we consider the vector $g_2'$ as 
\begin{equation}\label{eqn112}
    g_2'=\left(f+\frac{q}{2}z_{i_{\pi(1)}}+c_0\right)|_{\mathbf{z}z_{i_{\pi(0)}}=\mathbf{c}0}~ .
\end{equation}
Substituting $\mathbf{z=c}$ and $z_{i_{\pi(0)}}=0$ in (\ref{eqn112}) the function
   $ h_1+\frac{q}{2}z_{i_{\pi(1)}}+c_0$ is obtained
 which is identical to $h_2$. In component $i$, the value of the vector $g_2$ is the same as the value of the vector $g_2'$ in the position $i-2^{z_{i_{\pi(0)}}}$ (i.e., in non-zero positions, $g_2$ is simply a shift of $g_2'$). Therefore, the vectors $g_2$ and $g_2'$ have identical AACFs.
%It follows that the value of the vector $g_2$ in component $i$ is equal to that of the vector $g_2'$ in position $i-2^{x_{i_{\pi(0)}}}$ (i.e. in the non-zero positions, g_2 is just a shift of $g_2'$). So, the vectors $g_2$ and $g_2'$ have identical AACF.
Now, consider the pair
\begin{equation}
    g_1=f|_{\mathbf{z}z_{i_{\pi(0)}}=\mathbf{c}0},
\end{equation}
and
\begin{equation}
    g_2'=\left(f+\frac{q}{2}z_{i_{\pi(1)}}+c_0\right)|_{\mathbf{z}z_{i_{\pi(0)}}=\mathbf{c}0}.
\end{equation}
From the above, it is observed that $g_1$ corresponds to a GBF $h_1$ such that the graph of the quadratic part of $h_1$ is a path on $m-k-5$ vertices. Additionally, either $i_{\pi(1)}$ is an end vertex of this path, or $k=m-6$ and it is the single vertex in the graph. By the inductive hypothesis, $g_1$ and $g_2'$ forms a GCP, hence for $s \neq 0$, the sum of AACF of $g_1$ and $g_2'$ is
\begin{equation}
    \mathcal{A}(g_1)(s)+\mathcal{A}(g_2')(s)=0.
\end{equation}
Since, $\mathcal{A}(g_2)(s)=\mathcal{A}(g_2')(s)$ for every $s$, the sum of AACF  $g_1$ and $g_2$ is expressed as
\begin{equation}\label{eqn12}
  \mathcal{A}(g_1)(s)+\mathcal{A}(g_2)(s)=0.
\end{equation}
From the definitions, we have $g_3=\omega^{c'} g_1$ and $g_4=-\omega^{c'}g_2 .$ It follows that $\mathcal{A}(g_3)(s)=\mathcal{A}(g_1)(s)$ and $\mathcal{A}(g_4)(s)=\mathcal{A}(g_2)(s)$, so from (\ref{eqn12}), the sum of AACF  $g_3$ and $g_4$ is
\begin{equation}\label{eqn13}
      \mathcal{A}(g_3)(s)+\mathcal{A}(g_4)(s)=0. 
\end{equation}
Also, the ACCF between $g_3$ and $g_4$ is defined as
\begin{equation}
\begin{aligned}
    \mathcal{C}(g_3,g_4)(s)&=\mathcal{C}(\omega^{c'}g_1,-\omega^{c'}g_2)(s)\\& =-\mathcal{C}(g_1,g_2)(s).
\end{aligned}
\end{equation}
So the sum of ACCFs of $g_1,g_2$ and $g_3,g_4$ is 
\begin{equation}\label{eqn14}
\begin{aligned}
    \mathcal{C}(g_1,g_2)(s)&+\mathcal{C}(g_3,g_4)(s)\\& =\mathcal{C}(g_2,g_1)(s)+\mathcal{C}(g_4,g_3)(s)=0.
\end{aligned}
\end{equation}
So, from (\ref{eqn12})-(\ref{eqn14}), the sum in (\ref{eqn11}) is zero. Since $s\neq0$ has been chosen arbitrary, it follows that $f|_\mathbf{z=c}$ and $\left(f+\frac{q}{2}z_{\beta_2}+c'\right)|_\mathbf{z=c}$ forms a GCP of length $2^{m-1}+2^{m-3}$.
\end{IEEEproof}
\begin{example}\label{ex1}
For $m=8$ and $q=2$, consider the $5$th order GBF $f:\{\mathbf{0}_8,\mathbf{1}_8,\hdots,\mathbf{159}_8\} \rightarrow \mathbb{Z}_2$ defined as
\begin{equation}
    \begin{aligned}
        f=&z_0z_1+z_1z_2+z_2z_3+z_3z_0+z_0z_2+z_1z_3+z_0+z_1\\&+z_2+z_3+\bar z_7\left(\bar z_4\left(z_5+z_6\right)+z_6z_5\right)+z_2\left(\bar z_7\left(z_6\bar z_5\bar z_4 \right. \right.\\& \left. \left.+z_6z_5\right)+z_7\bar z_6\bar z_5\right).
    \end{aligned}
\end{equation}
\begin{figure}[h]
    \centering
    \includegraphics{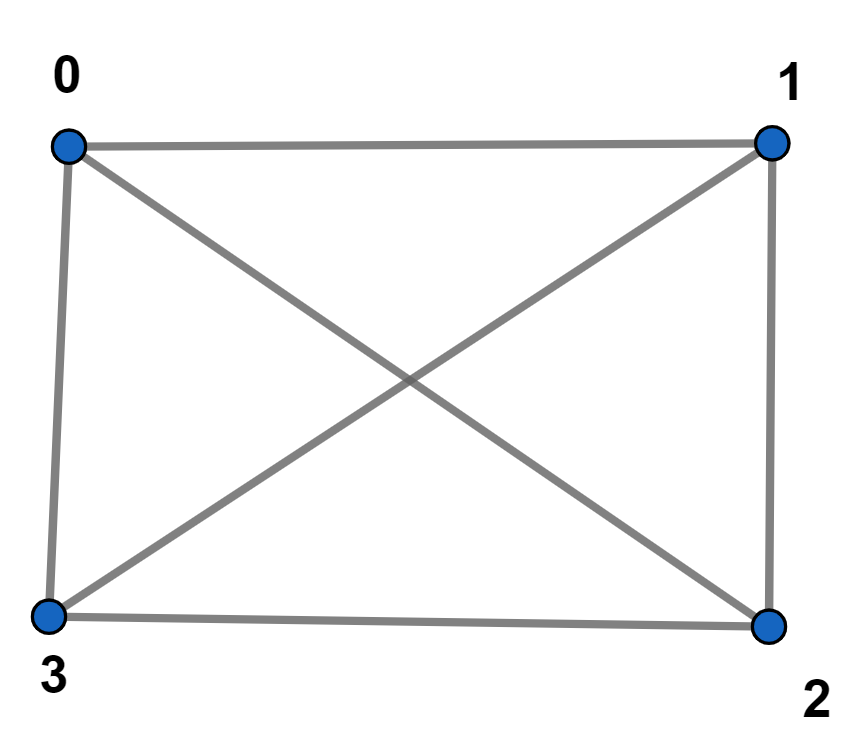}
    \caption{The graph of quadratic part $Q$ of $f$}
    \label{fig:my_label}
\end{figure}
The graph $G(Q)$ (quadratic part of $f$) is given in Fig. \ref{fig:my_label}. By substituting $z_0z_3=00$ ( deleting vertices $z_0,z_3$), we get $G\left(Q|_{z_0z_3=00}\right)$ is a path. So by \textit{Theorem} \ref{thm1}, $f|_{z_0z_3=00}$ and $\left(f+z_1+1\right)|_{z_0z_3=00}$ forms a GCP of length 160, which is not the form of $2^m$.
\end{example}
\section{Proposed Construction of MOCSs}\label{sec4}
In this section, we have proposed a direct construction of $2^k$ CSs of length $2^{m-1}+2^{m-3}$, with the property that any two CSs are mutually orthogonal to each other. 

Let $Q$ and $f$ be defined in (\ref{eqn5}) and (\ref{eqn7}) respectively ($q=2$). For $0\leq t <2^k$, $0\leq k\leq m-5$, the ordered set $S_t$ (with the natural order induced by the binary vector $(aa_0a_1\cdots a_{k-1})$) is defined as
\begin{equation}\label{eqn15}
   S_t= \left\{f+\sum_{\alpha=0}^{k-1} a_{\alpha} z_{p_{\alpha}}+\sum_{\alpha=0}^{k-1} n_{\alpha} z_{p_{\alpha}}+a z_{\beta_2}: a,a_\alpha \in \{0,1\}\right\},
\end{equation}
where $t=\sum_{\alpha=0}^{k-1} n_{\alpha} 2^{\alpha}$.\\
%\textcolor{red}
``$\mathbf{1}$" represents a vector all of whose component is one and $\oplus$ denotes addition modulo 2.
\begin{theorem}\label{thm2}
Suppose that $G(Q)$ contains a set of $k \leq m-5$ distinct vertices labeled $p_0,p_1,\hdots,p_{k-1}$ with the property that deleting those $k$ vertices and all their edges results in a path. Let $\beta_1$ and $\beta_2$ be the two end vertices of the path. In case of single vertex let $\beta_1=\beta_2=\beta$. Then %for any choice of $c, c_{i} \in \mathbb{Z}_{q}$,
for any  $0 \leq t<2^{k}$,
the set $S_{t}$ %(for $L=2^{m-2}+2^{m-3}$)
is a CS. Also for the case $t' \neq t$, the sets $S_{t'}$ and $S_{t}$ are MOCSs.
\end{theorem}
\begin{IEEEproof}
Since each $S_t$ for $1\leq t <2^k$ is a permutation of $S_0$, so  proving $S_0$ is a complementary set is sufficient to show that for any $0\leq t <2^k$, the set $S_t$ is a CS.\\ 
Let $\mathbf{z}=\left(z_{p_0}z_{p_1}\hdots z_{p_{k-1}}\right)$ and $\mathbf{a}=\left(a_{0}a_{1}\hdots a_{k-1}\right)$. So $\mathbf{a\cdot z}=\sum_{\alpha=0}^{k-1}a_\alpha z_{p_\alpha}$. Now from \textit{Lemma} \ref{lem2}, for $s \neq 0$, sum of AACF can be expressed as 
\begin{equation}
    \sum_{\mathbf{a}, a} \mathcal{A}\left(f+\mathbf{a \cdot z}+a z_{\beta_2}\right)(s)=L_{1}+L_{2},
\end{equation}
where
\begin{equation}\label{eqn16}
L_{1}=\sum_{\mathbf{a}, a} \sum_{\mathbf{c}} \mathcal{A}\left(\left(f+\mathbf{a \cdot z}+a z_{\beta_{2}}\right)|_{\mathbf{z}=\mathbf{c}}\right)(s),
\end{equation}
and
\begin{equation}\label{eqn17}
\begin{aligned}
L_{2}=\sum_{\mathbf{c_{1} \neq c_{2}}}\sum_{a} \sum_{\mathbf{a}} &C\left(\left(f+\mathbf{a \cdot z}+a z_{\beta_{2}}\right)|_{\mathbf{z}=\mathbf{c}_{1}},\right.\\& \left. 
\left(f+\mathbf{a \cdot z}+a z_{\beta_{2}}\right)|_{\mathbf{z=c_2}}\right)(s).
\end{aligned}
\end{equation}
 The graph of the  function $\left(Q+\mathbf{a \cdot z}\right)|_\mathbf{z=c}$ is a path for any choice of $\mathbf{c}$ and $\mathbf{a}$. So from \textit{Theorem} \ref{thm1}, for every $\mathbf{c}$ and $\mathbf{a}$ the vectors $\left(f+\mathbf{a \cdot z}\right)|_\mathbf{z=c}$ and $\left(f+\mathbf{a \cdot z}+z_{\beta_{2}}\right)|_\mathbf{z=c}$ forms a GCP of length $2^{m-1}+2^{m-3}$. Hence the term $L_1$ in (\ref{eqn16}) is zero.
For fixed values of $\mathbf{c_1,c_2}$ and $a$, consider the inner sum of $L_2$ is expressed as
\begin{equation}\label{eqn18}
\sum_{\mathbf{a}} \mathcal{C}\left(\left(f+\mathbf{a \cdot z}+a z_{\beta_{2}}\right)|_{\mathbf{z}=\mathbf{c}_{1}}, 
\left(f+\mathbf{a \cdot z}+a z_{\beta_{2}}\right)|_{\mathbf{z=c_2}}\right)(s).
\end{equation}
Now, the vector $\mathbf{z}$ contains all the terms of $\mathbf{a \cdot z}$. So for the fixed values of $\mathbf{c_1,c_2}$ and $a$ we have,
\begin{equation}\label{eqn120}
\begin{aligned}
\left(f \right.& \left.+\mathbf{a \cdot z}+a  z_{\beta_{2}}\right)|_{\mathbf{z}=\mathbf{c}_j}\\&=
    \begin{cases}
   \mathbf{e_j}= \left(f+az_{\beta_{2}}\right)|_\mathbf{z=c_j},&\text {when}~\mathbf{a\cdot c_j}=0 \Mod2,\\
   -\mathbf{e_j},&\text {when}~\mathbf{a\cdot c_j}=1 \Mod2.
    \end{cases}
    \end{aligned}
\end{equation}
Therefore for the fixed values of $\mathbf{c_1,c_2}$ and $a$, from (\ref{eqn18}) and (\ref{eqn120}) ACCF values are obtained as
\begin{equation}
\begin{aligned}
& \mathcal{C}\left(\left(f+\mathbf{a \cdot z}+a z_{\beta_{2}}\right)|_{\mathbf{z}=\mathbf{c}_{1}}, 
\left(f+\mathbf{a \cdot z}+a z_{\beta_{2}}\right)|_{\mathbf{z=c_2}}\right)(s)
\\&\!=\!
    \begin{cases}
    \mathcal{C}(\mathbf{e_1,e_2}),&\!\text {when}~\mathbf{a\cdot c_1=a\cdot c_2}=0 \Mod2,\\
   \mathcal{C}(\mathbf{\!-e_1,\!-e_2\!}),&\!\text {when}~\mathbf{a\cdot c_1=a\cdot c_2}=1 \Mod2,\\
    \mathcal{C}(\mathbf{e_1,-e_2}),&\!\text {when}~\mathbf{\!a\!\cdot\! c_1\!}\!=\!0\!\Mod2, \mathbf{\!a\!\cdot\! c_2}\!=\!1\! \Mod2,\\
    \mathcal{C}(\mathbf{-e_1,e_2}),&\!\text {when}~\mathbf{\!a\!\cdot\! c_1\!}\!=\!1\! \Mod2, \mathbf{\!a\!\cdot\! c_2\!}\!=\!0\!\Mod2.\\
    \end{cases}
    \end{aligned}
\end{equation}
Since,  $\mathcal{C}(\mathbf{e_1,e_2})= \mathcal{C}(\mathbf{-e_1,-e_2})$ and  $\mathcal{C}(\mathbf{e_1,-e_2})= \mathcal{C}(\mathbf{-e_1,e_2})$, the above can be re-expressed as
\begin{equation}
\begin{aligned}
& \mathcal{C}\left(\left(f+\mathbf{a \cdot z}+a z_{\beta_{2}}\right)|_{\mathbf{z}=\mathbf{c}_{1}},
\left(f+\mathbf{a \cdot z}+a z_{\beta_{2}}\right)|_{\mathbf{z=c_2}}\right)(s)
\\&=
    \begin{cases}
    \mathcal{C}(\mathbf{e_1,e_2}),&\text {when}~\mathbf{a \cdot c_1=a\cdot c_2} \Mod2,\\
    \mathcal{C}(\mathbf{e_1,-e_2}),&\text {when}~\mathbf{a \cdot c_1} \neq \mathbf{a\cdot c_2}\Mod2.\\
    \end{cases}
    \end{aligned}
\end{equation}
\begin{equation}\label{eqn119}
\begin{aligned}
&\sum_{\mathbf{a}} \mathcal{C}\left(\left(f+\mathbf{a \cdot z}+a z_{\beta_{2}}\right)|_{\mathbf{z}=\mathbf{c}_{1}}, 
\left(f+\mathbf{a \cdot z}+a z_{\beta_{2}}\right)|_{\mathbf{z=c_2}}\right)(s)\\&\!=\!
\sum_{\mathbf{a\cdot c_1}\!= \mathbf{a \cdot c_2\!}\!\Mod 2\!} \!\mathcal{C}(\mathbf{e_1,e_2})(s)\! - \!\sum_{\mathbf{a\cdot c_1}\neq \mathbf{a \cdot c_2\!}\!\Mod 2\!} \!\mathcal{C}(\mathbf{e_1,e_2})(s).
\end{aligned}
\end{equation}
Due to the fact that $\mathbf{c_1}\neq \mathbf{c_2}$, $\mathbf{c_1}+\mathbf{c_2}\neq \mathbf{0} \Mod2$, and so the linear functional $\mathbf{a} \cdot  \left(\mathbf{c}_{1}+\mathbf{c}_{2}\right) \Mod 2$  takes each value $0$ and $1$ precisely $2^{k-1}$ times, i.e., an equal number of times.
% Since $\mathbf{c_1}\neq \mathbf{c_2}$, it implies that $\mathbf{c_1}+\mathbf{c_2}\neq \mathbf{0} \Mod2$ 
%and so the linear functional $\mathbf{a} \cdot  %\left(\mathbf{c}_{1}+\mathbf{c}_{2}\right) \Mod 2$ takes each value $0$ and $1$ at exactly $2^{k-1}$ times, i.e., equal number of times.
So, from (\ref{eqn119}) the inner sum of $L_2$ is zero and so is $L_2$. Hence it is proved that $S_t$ is a CS of size $2^k$. \\
Now, let $\mathbf{n}=(n_{0} n_{1} \cdots n_{k-1})$,
$t=\sum_{\alpha=0}^{k-1} n_{\alpha} 2^{\alpha}$ and $t'=\sum_{\alpha=0}^{k-1} n'_{\alpha} 2^{\alpha}$. It needs to proven that for $t\neq t'$, $S_t$ and
$S_{t'}$ are mutually orthogonal. From \textit{Lemma} \ref{lem1}, the sum of ACCF can be written as 
\begin{equation}
\begin{aligned}
\sum_{\mathbf{a}, a} &\mathcal{C}\left(f+ \left(\mathbf{a}+\mathbf{n}\right) \cdot \mathbf{z}+a z_{\beta_{2}},\right.\\& \left.
 f+\left(\mathbf{a}+\mathbf{n}^{\prime}\right) \cdot \mathbf{z}+a z_{\beta_{2}}\right)(s)=M_1+M_2,
\end{aligned}
\end{equation}
where
\begin{equation}\label{eqn19}   
\begin{aligned}
M_{1}=\sum_{\mathbf{a}, a} \sum_{\mathbf{c}_{1} \neq \mathbf{c}_{2}} &\mathcal{C}\left(\left(f+\left(\mathbf{a}+\mathbf{n}\right) \cdot \mathbf{z}+a z_{\beta_{2}}\right)|_{\mathbf{z}=\mathbf{c}_{1}}, \right. \\& \left.
\left(f+\left(\mathbf{a}+\mathbf{n}^{\prime}\right) \cdot \mathbf{z}+a z_{\beta_{2}}\right)|_{\mathbf{z}=\mathbf{c}_{2}}\right)(s),
\end{aligned}
\end{equation}
and
\begin{equation} \label{eqn20}   
\begin{aligned}
M_{2}=\sum_{\mathbf{a}, a} \sum_{\mathbf{c}} &\mathcal{C}\left(\left(f+\left(\mathbf{a}+\mathbf{n}\right) \cdot \mathbf{z}+a z_{\beta_{2}}\right)|_{\mathbf{z}=\mathbf{c},} \right. \\& \left.
\left(f+\left(\mathbf{a}+\mathbf{n}^{\prime}\right) \cdot \mathbf{z}+a z_{\beta_{2}}\right)|_{\mathbf{z}=\mathbf{c}}\right)(s).
\end{aligned}
\end{equation}
For the fixed $\mathbf{c_1,c_2}$ and $a$, we consider the following sum of $M_1$
\begin{equation*}    
\begin{aligned}
\sum_{\mathbf{a}} &\mathcal{C}\left(\left( f+\left(\mathbf{a}+\mathbf{n}\right) \cdot \mathbf{z}+a z_{\beta_{2}}\right)|_{\mathbf{z}=\mathbf{c}_{1}}, \right. \\& \left.
\left(f+\left(\mathbf{a}+\mathbf{n}^{\prime}\right) \cdot \mathbf{z}+a z_{\beta_{2}}\right)|_{\mathbf{z}=\mathbf{c}_{2}}\right)(s)
\end{aligned}
 \end{equation*} 
 \begin{equation*}
     \begin{aligned}
~~~~=&\sum_{\mathbf{a}} \mathcal{C}\left(\left(f+\left(\mathbf{a}+\mathbf{n}\right) \cdot \mathbf{c_1}+a z_{\beta_{2}}\right)|_{\mathbf{z}=\mathbf{c}_{1}},\right. \\& \left.
\left(f+\left(\mathbf{a}+\mathbf{n}^{\prime}\right) \cdot \mathbf{c_2}+a z_{\beta_{2}}\right)|_{\mathbf{z}=\mathbf{c}_{2}}\right)(s)
\end{aligned}
\end{equation*}
\begin{equation*}
    \begin{aligned}
~~~~~~~~~~~~=&\sum_{\mathbf{a}}(-1)^{\mathbf{a \cdot(c_1\oplus  c_2)}} \mathcal{C}\left(\left(f+\mathbf{n} \cdot \mathbf{z}+a z_{\beta_{2}}\right)|_{\mathbf{z}=\mathbf{c}_{1}}, \right. \\& \left.
\left(f+\mathbf{n}^{\prime} \cdot \mathbf{z}+a z_{\beta_{2}}\right)|_{\mathbf{z}=\mathbf{c}_{2}}\right)(s)
\end{aligned}
\end{equation*}
\begin{equation}\label{eqn21}
    \begin{aligned}
~~~~~~~~~~~~~=& \mathcal{C}\left(\left(f+\mathbf{n} \cdot \mathbf{z}+a z_{\beta_{2}}\right)|_{\mathbf{z}=\mathbf{c}_{1}}, \right. \\& \left.
\left(f+\mathbf{n}^{\prime} \cdot \mathbf{z}+a z_{\beta_{2}}\right)|_{\mathbf{z}=\mathbf{c}_{2}}\right)(s)\sum_{\mathbf{a}}(-1)^{\mathbf{a \cdot (c_1\oplus c_2)}}.
\end{aligned}
\end{equation}
Since $\mathbf{c_1}\neq \mathbf{c_2}$, so the function $\mathbf{a \cdot(c_1\oplus c_2)}$ in (\ref{eqn21}) takes values $0$ and $1$ equal number of times and hence (\ref{eqn21}) vanishes for all $s$.

Now for the fixed $\mathbf{a}$ and $\mathbf{c}$  consider the following sum of $M_2$
\begin{equation*}    
\begin{aligned}
 \sum_{{a}} &\mathcal{C}\left(\left(f+\left(\mathbf{a}+\mathbf{n}\right) \cdot \mathbf{z}+a z_{\beta_{2}}\right)|_{\mathbf{z}=\mathbf{c},} \right. \\& \left.
\left(f+\left(\mathbf{a}+\mathbf{n}^{\prime}\right) \cdot \mathbf{z}+a z_{\beta_{2}}\right)|_{\mathbf{z}=\mathbf{c}}\right)(s)
\end{aligned}
\end{equation*}
\begin{equation*}
     \begin{aligned}
~~~~=&\sum_{{a}} \mathcal{C}\left(\left(f+\left(\mathbf{a}+\mathbf{n}\right) \cdot \mathbf{c}+a z_{\beta_{2}}\right)|_{\mathbf{z}=\mathbf{c}},\right. \\& \left.
\left(f+\left(\mathbf{a}+\mathbf{n}^{\prime}\right) \cdot \mathbf{c}+a z_{\beta_{2}}\right)|_{\mathbf{z}=\mathbf{c}}\right)(s)
\end{aligned}
\end{equation*}
\begin{equation*}
     \begin{aligned}
~~~~=&\sum_{{a}} \mathcal{C}\left(\left(f+\left(\mathbf{n}+\mathbf{n'}\right) \cdot \mathbf{c}+a z_{\beta_{2}}\right)|_{\mathbf{z}=\mathbf{c}},\right. \\& \left.
\left(f+a z_{\beta_{2}}\right)|_{\mathbf{z}=\mathbf{c}}\right)(s)
\end{aligned}
\end{equation*}
\begin{equation*}
=(-1)^{\mathbf{(n\oplus n')}\cdot c}\sum_{{a}} \mathcal{C}\left(\left(f+a z_{\beta_{2}}\right)|_{\mathbf{z}=\mathbf{c}},%\right. \left.
\left(f+a z_{\beta_{2}}\right)|_{\mathbf{z}=\mathbf{c}}\right)(s)
\end{equation*}
\begin{equation}\label{eqn22}
=(-1)^{\mathbf{(n\oplus n')}\cdot c}\mathcal{A}\left(f|_\mathbf{z=c}\right)(s)%\\& \left.
%\right.
+\mathcal{A}\left(\left(f+ z_{\beta_{2}}\right)|_{\mathbf{z}=\mathbf{c}}\right) (s).
\end{equation}
From \textit{Theorem} \ref{thm1}, the above sum in (\ref{eqn22}) is zero for all $s \neq 0$. For $s=0$, AACF is given by
\begin{equation}\label{eqn23}
    \mathcal{A}\left(f|_\mathbf{z=c}\right)(s)=\mathcal{A}\left(\left(f+ z_{\beta_{2}}\right)|_{\mathbf{z}=\mathbf{c}}\right) (s)=2^{m-k-4},
\end{equation}
for $\mathbf{c} \in \mathbb{Z}_2^k$, substituting this back in (\ref{eqn22}), we get the sum of ACCF as
\begin{equation}\label{eqn24}
     \begin{aligned}
&\sum_{{a}} \mathcal{C}\left(\left(f+\left(\mathbf{n}+\mathbf{n'}\right) \cdot \mathbf{c}+a z_{\beta_{2}}\right)|_{\mathbf{z}=\mathbf{c}},\right. \\& \left.
\left(f+a z_{\beta_{2}}\right)|_{\mathbf{z}=\mathbf{c}}\right)(0)=(-1)^{\mathbf{(n\oplus n')}\cdot c}\cdot 2^{m-k-3}.
\end{aligned}
\end{equation}
Here $t\neq t'$ is considered, which implies  $\mathbf{n} \neq \mathbf{n}^{\prime}$, and hence $\mathbf{n} \oplus \mathbf{n}^{\prime} \neq 0$. So the linear functional $\left(\mathbf{n} \oplus \mathbf{n}^{\prime}\right) \cdot \mathbf{c}$ (regarded as a function of $\left.\mathbf{c}\right)$ is not equivalent to the zero function. As a result, it is balanced, i.e., the values 0 and 1 are taken equal number of times by the function as $\mathbf{c}$ varies. Hence the sum
\begin{equation}
\sum_{\mathbf{c}}(-1)^{\left(\mathbf{n} \oplus \mathbf{n}^{\prime}\right) \cdot \mathbf{c}} \cdot 2^{m-k+1}=0.
\end{equation}
%\vspace{-0.2cm}
\end{IEEEproof}
\begin{remark}
\cite[Th. 4]{chen1} generates CSs of length $2^{m-1}+2^{m-3}$ and set size $4$ for $\nu=m-3$, which is covered by \textit{Theorem} \ref{thm2} of our proposed construction by taking $k=2$.
\end{remark}
\begin{remark}
By taking $\nu=m-3$,  \cite[Th. 4]{chen2} and $t=m-3$, \cite[Th. 3]{linyu} generates CSs of length $2^{m-1}+2^{m-3}$ and set size $2^{k+1}$. The proposed construction of CSs in \textit{Theorem} \ref{thm2}  covers these special cases of \cite{chen1,linyu} .
\end{remark}
\textit{TABLE} \ref{table 1} compares the proposed constructions of MOCSs with the existing direct constructions of \cite{shing,liying}.
 \begin{table}[ht]
%\tiny
\centering
\caption{Comparison of the proposed MOCS construction with \cite{shing,liying}}
\resizebox{\textwidth}{!}{
\begin{tabular}{|l|l|l|l|l|}
\hline
     Ref. & Parameters & Based on   & Length(N) & Constraint %& \makecell{Column sequence\\ PMEPR is upper\\ bounded by }                                       
     \\ \hline
        \cite{shing}       & $(2^{k'},2^{k+1},N)$  & GBF of order $ 2 $   &$2^{m-1}+2^{t}$           & \makecell{$m,k,t\in \mathbb{Z}^+$, $m\geq 2,k\leq m$,\\ $ 0\leq k'\leq t\leq m-1,k'\leq k-1$ }                 \\ \hline
        
        \cite{liying}       &$(2^k,2^{k+1},N)$ &   GBF of order $2$  & $2^m+2^t$  & $m,k,t \in \mathbb{Z}^+$, $0\leq t <k \leq m $                            \\ \hline
    %\cite{zhou} & $(K_2,K_1K_2,)$ & \makecell{$(K_2,K_2,L_2)$-CCC \\ $(K_1,L_1)$-even shift CS} &  $L_1L_2$ &  & $MN$ \\ \hline
    
\multirow{2}{*}{\textit{Theorem} \ref{thm2}}   & \multirow{2}{*}{$(2^{k+1},2^{k+1},N)$} & \multirow{2}{*}{GBF of order $> 2$ }    &$2^{m-1}+2^{m-3}$        &$m,k\in \mathbb{Z}^+,m\geq 5$   \\
\cline{4-5} &  & & $2^{m-1}+2^{m-2}+2^{m-4}$ & $m,k\in \mathbb{Z}^+,m\geq 6$
\\ \hline
\end{tabular}}\label{table 1}
\end{table}
\begin{example}\label{ex2}
Let us consider the same GBF as given in \textit{Example} \ref{ex1}, and the deleted vertices are also same, i.e., $z_0,z_3$. Then the set,
\begin{equation}
    \begin{aligned}
       S_0=&\left\{f,f+z_1,f+z_0,f+z_0+z_1,f+z_3\right.\\& \left.,f+z_3+z_1,f+z_3+z_0,f+z_3+z_1+z_0 \right\},
     \end{aligned}
\end{equation}
is a CS of size 8 and sequence length 160, which is not of the form of $2^m$. Similarly the sets $S_t$ for $0\leq t < 4$, which are the permutations of the set $S_0$, are also CS of size 8, with the property that any two different CSs are mutually orthogonal to each other. 
\end{example}
 \section{Proposed Construction of CCCs}\label{sec5}
 In this section first we construct a mate of the MOCSs proposed in section \ref{sec4}. Then binary CCCs of length $2^{m-1}+2^{m-3}$ are constructed by union of these two MOCSs through GBFs. 
 For a given GBF  $f$ in (\ref{eqn7}), GBF $\bar f:B \rightarrow \mathbb{Z}_2$ is defined as 
\begin{equation}\label{eqn52}
    \bar f(z_0,z_1,\hdots,z_{m-1})=f(\bar z_0,\bar z_1,\hdots,\bar z_{m-1}),
\end{equation}
 where $B=\{0,1\}^m \setminus \left\{\mathbf{0}_m,\mathbf{1}_m,\hdots,(\mathbf{2^{m-2}+2^{m-3}-1})_m \right\}$.
\begin{lemma}\label{lem3}
%Suppose that $G(Q)$ contains a set of $k \leq m-5$ distinct vertices labeled with the property that deleting those $k$ vertices and all their edges results in a path.
Let us assume a set of $k \leq m-5$ distinct vertices labelled with the property that deleting that set of vertices and all the edges transform $G(Q)$ into a path.
Let $\beta_1$ and $\beta_2$ be the two end vertices of this path. In case of $k=m-5$, the single vertex of the graph is denoted by $\beta_1=\beta_2=\beta$. Then for each $0\leq t <2^k$, the ordered set $\bar S_t$ %(for $L=2^{m-2}+2^{m-3}$)
given by
%\vspace{-0.02cm}
\begin{equation}\label{eqn25}
    \left\{\bar f+\sum_{\alpha=0}^{k-1} a_{\alpha} \bar z_{p_{\alpha}}+\sum_{\alpha=0}^{k-1} n_{\alpha} \bar z_{p_{\alpha}}+\bar a z_{\beta_{2}}: a,a_\alpha \in \{0,1\}\right\},
\end{equation}
is a CS of size $2^{k+1}$, where $\bar f$  is defined in (\ref{eqn52}). %$t=\sum_{\alpha=0}^{k-1} n_{\alpha} 2^{\alpha}$.
Further, for $t'\neq t$, $\bar S_{t'}$ and $\bar S_{t}$ are MOCSs, where the natural order is induced from the binary vector $(aa_0a_1\cdots a_{k-1})$.
\end{lemma}
The next theorem gives CCCs of length $2^{m-1}+2^{m-3}$. 
\begin{theorem}\label{thm3}
Let the sets $S_t$ and $\bar S_t$ be defined in \textit{Theorem} \ref{thm2} and \textit{Lemma} \ref{lem3} respectively, then 
%\vspace{-0.1cm}
\begin{equation}\label{eqn26}
    \left\{S_t: 0\leq t<2^k\right\}
    \cup\left\{\bar S_t: 0\leq t<2^k\right\},
\end{equation} %\vspace{-1cm}
forms a $\left(2^{k+1},2^{k+1},2^{m-1}+2^{m-3}\right)$-CCC.
\end{theorem}
\begin{IEEEproof}
It will be shown that CSs $S_{t_{1}}$ and $\bar S_{t_{2}}$ are mutually orthogonal to each other. The sum of ACCF of these CSs can be expressed as  
\begin{equation}\label{eqn27}
    \begin{aligned}
\sum_{\mathbf{a}} &\mathcal{C}\left( f+\left(\mathbf{a}+\mathbf{n}\right) \cdot \mathbf{z}+ z_{\beta_{2}},
\bar f+\left(\mathbf{a}+\mathbf{n}^{\prime}\right) \cdot \mathbf{\bar z}\right)(s) \\&  +\mathcal{C}\left( f+\left(\mathbf{a}+\mathbf{n}\right) \cdot \mathbf{z}\right),
\left(\bar f+\left(\mathbf{a}+\mathbf{n}^{\prime}\right) \cdot \mathbf{\bar z}+z_{\beta_{2}}\right)(s)\\
=&\sum_{\mathbf{a}}\sum_{\mathbf{c_1,c_2}} \mathcal{C}\left(\left( f+\left(\mathbf{a}+\mathbf{n}\right) \cdot \mathbf{z}+ z_{\beta_{2}}\right)|_\mathbf{z=c_1},\right. \\& \left.
\hspace{2.5cm}\left(\bar f+\left(\mathbf{a}+\mathbf{n}^{\prime}\right) \cdot \mathbf{\bar z}\right)|_\mathbf{z=c_2} \right)(s) \\&  +\mathcal{C}\left(\left( f+\left(\mathbf{a}+\mathbf{n}\right) \cdot \mathbf{z}\right)|_\mathbf{z=c_1},\right.\\& \left.
\hspace{1cm}\left(\bar f+\left(\mathbf{a}+\mathbf{n}^{\prime}\right) \cdot \mathbf{\bar z}+z_{\beta_{2}}\right)|_\mathbf{z=c_2}\right)(s) = M(say)
\end{aligned}
\end{equation}
For a given $\mathbf{c_1}$ and $\mathbf{c_2}$, consider the following sum of the first term in (\ref{eqn27}) 
%\vspace{-0.8cm}
\begin{equation}\label{eqn28}
    \begin{aligned}
\sum_{\mathbf{a}} &\mathcal{C}\left(\left( f+\left(\mathbf{a}+\mathbf{n}\right) \cdot \mathbf{z}+ z_{\beta_{2}}\right)|_\mathbf{z=c_1},\right. \\&~~ \left.
\left(\bar f+\left(\mathbf{a}+\mathbf{n}^{\prime}\right) \cdot \mathbf{\bar z}\right)|_\mathbf{z=c_2} \right)(s) \\&
=\sum_{\mathbf{a}} \mathcal{C}\left(\left( f+\left(\mathbf{a}+\mathbf{n}\right) \cdot \mathbf{z}+ z_{\beta_{2}}\right)|_\mathbf{z=c_1},\right. \\& \left.
~~\left(\bar f+\left(\mathbf{a}+\mathbf{n}^{\prime}\right) \cdot \mathbf{(1-z)}\right)|_\mathbf{z=c_2} \right)(s)\\&
= \mathcal{C}\left(\left( f+ z_{\beta_{2}}\right)|_\mathbf{z=c_1},
(\bar f|_\mathbf{z=c_2} \right)(s)\\&
\cdot \sum_{\mathbf{a}}\left((-1)^{\mathbf{n} \cdot \mathbf{c}_{1} \oplus \mathbf{n}^{\prime} \cdot \mathbf{c}_{2}} \cdot(-1)^{\left(\mathbf{a} \oplus \mathbf{n}^{\prime}\right) \cdot \mathbf{1}} \cdot(-1)^{\mathbf{a} \cdot\left(\mathbf{c}_{1} \oplus \mathbf{c}_{2}\right)}\right)\\&
= \mathcal{C}\left(\left( f+ z_{\beta_{2}}\right)|_\mathbf{z=c_1},
(\bar f|_\mathbf{z=c_2} \right)(s)\\&
~~\cdot (-1)^{\mathbf{n} \cdot \mathbf{c}_{1} \oplus \mathbf{n}^{\prime} \cdot \mathbf{c}_{2} \oplus \mathbf{n'}\cdot \mathbf{1}}\sum_{\mathbf{a}}(-1)^{\mathbf{a} \cdot\left(\mathbf{c}_{1} \oplus \mathbf{c}_{2} \oplus \mathbf{1}\right)}.
\end{aligned}
\end{equation}
The above sum in (\ref{eqn28}) vanishes whenever $\left(\mathbf{c}_{1} \oplus \mathbf{c}_{2}\right) \neq \mathbf{1}$. So, the first correlation term in (\ref{eqn27}) is zero whenever $\mathbf{c_1}$ and $\mathbf{c_2}$ are equal. Thus, summing (\ref{eqn28}) over all $\mathbf{c_1} \neq \mathbf{c_2}$, the above term further can be simplified as
\begin{equation}\label{eqn29}
    \begin{aligned}
    &\sum_{\mathbf{c_1}\neq \mathbf{c_2}\atop \mathbf{c_1+c_2=1}}\mathcal{C}\left(\left( f+ z_{\beta_{2}}\right)|_\mathbf{z=c_1},
(\bar f|_\mathbf{z=c_2} \right)(s)\\&
\hspace{1.5cm}
\cdot (-1)^{\mathbf{n} \cdot \mathbf{c}_{1} \oplus \mathbf{n}^{\prime} \cdot \mathbf{c}_{2} \oplus \mathbf{n'}\cdot \mathbf{1}}2^k \\&
=\sum_{\mathbf{c}}\mathcal{C}\left(\left( f+ z_{\beta_{2}}\right)|_\mathbf{z=c},
(\bar f|_{\mathbf{z=(c\oplus1})} \right)(s)\\&
\hspace{1.5cm}
\cdot(-1)^{\mathbf{n'}\cdot \mathbf{1}}2^k \cdot (-1)^{\mathbf{n} \cdot \mathbf{c} \oplus \mathbf{n}^{\prime} \cdot (1\oplus\mathbf{c})} \\&
=\sum_{\mathbf{c}}\mathcal{C}\left(\left( f+ z_{\beta_{2}}\right)|_\mathbf{z=c},
(\bar f|_{\mathbf{z=(c\oplus1})} \right)(s)
2^k \cdot (-1)^{\left(\mathbf{n} \oplus \mathbf{n}^{\prime}\right) \cdot \mathbf{c}}.
    \end{aligned}
\end{equation}
From \textit{Lemma} \ref{lem1}, the inner sum of  (\ref{eqn29}) can be further simplified as
\begin{equation*}
    \begin{aligned}
    &\mathcal{C}\left(\left( f+ z_{\beta_{2}}\right)|_\mathbf{z=c},
\bar f|_{\mathbf{z=(c\oplus1})} \right)(s)\\&
= \mathcal{C}\left(\left( f+ z_{\beta_{2}}\right)|_{\mathbf{z}z_{\beta_{2}}=\mathbf{c}0},
\bar f|_{\mathbf{z}z_{\beta_{2}}=(c\oplus1)0} \right)(s)\\&
~~+ \mathcal{C}\left(\left( f+ z_{\beta_{2}}\right)|_{\mathbf{z}z_{\beta_{2}}=\mathbf{c}0},
\bar f|_{\mathbf{z}z_{\beta_{2}}=(c\oplus1)1} \right)(s)\\&
~~+ \mathcal{C}\left(\left( f+ z_{\beta_{2}}\right)|_{\mathbf{z}z_{\beta_{2}}=\mathbf{c}1},
\bar f|_{\mathbf{z}z_{\beta_{2}}=(c\oplus1)0} \right)(s)\\&
~~+ \mathcal{C}\left(\left( f+ z_{\beta_{2}}\right)|_{\mathbf{z}z_{\beta_{2}}=\mathbf{c}1},
\bar f|_{\mathbf{z}z_{\beta_{2}}=(c\oplus1)1} \right)(s)\\&
\end{aligned}
\end{equation*}
\begin{equation}\label{eqn32}
\begin{aligned}
=&\mathcal{C}\left( f|_{\mathbf{z}z_{\beta_{2}}=\mathbf{c}0},
\bar f|_{\mathbf{z}z_{\beta_{2}}=(c\oplus1)0} \right)(s)\\&
~~+\mathcal{C}\left( f|_{\mathbf{z}z_{\beta_{2}}=\mathbf{c}0},
\bar f|_{\mathbf{z}z_{\beta_{2}}=(c\oplus1)1} \right)(s)\\&
~~-\mathcal{C}\left( f|_{\mathbf{z}z_{\beta_{2}}=\mathbf{c}1},
\bar f|_{\mathbf{z}z_{\beta_{2}}=(c\oplus1)0} \right)(s)\\&
~~-\mathcal{C}\left( f|_{\mathbf{z}z_{\beta_{2}}=\mathbf{c}1},
\bar f|_{\mathbf{z}z_{\beta_{2}}=(c\oplus1)1} \right)(s).
    \end{aligned}
\end{equation}
Similarly, the second term of the correlation in  (\ref{eqn27}) becomes
%\vspace{-0.3cm}
\begin{equation}\label{eqn30}
    \begin{aligned}
    &\sum_{\mathbf{d}} \mathcal{C}\left(\left( f+\left(\mathbf{d}+\mathbf{n}\right) \cdot \mathbf{z}\right)|_\mathbf{z=c_1},\right. \\& \left.
\hspace{1cm}\left(\bar f+\left(\mathbf{a}+\mathbf{n}^{\prime}\right) \cdot \mathbf{\bar z}+z_{\beta_{2}}\right)|_\mathbf{z=c_2} \right)(s) \\&
=\sum_{\mathbf{c}}\mathcal{C}\left( f|_\mathbf{z=c},
\left(\bar f+z_{\beta_{2}}\right)|_{\mathbf{z=(c\oplus1})} \right)(s)\cdot
2^k \cdot (-1)^{\left(\mathbf{n} \oplus \mathbf{n}^{\prime}\right) \cdot \mathbf{c}}.
    \end{aligned}
\end{equation}
The inner sum of (\ref{eqn30}) can be simplified as
\begin{equation}\label{eqn33}
    \begin{aligned}
  &\mathcal{C}\left( f|_\mathbf{z=c},
\left(\bar f+z_{\beta_{2}}\right)|_{\mathbf{z=(c\oplus1})} \right)(s) \\
=&\mathcal{C}\left( f|_{\mathbf{z}z_{\beta_{2}}=\mathbf{c}0},
\bar f|_{\mathbf{z}z_{\beta_{2}}=(c\oplus1)0} \right)(s)\\&
-\mathcal{C}\left( f|_{\mathbf{z}z_{\beta_{2}}=\mathbf{c}0},
\bar f|_{\mathbf{z}z_{\beta_{2}}=(c\oplus1)1} \right)(s)\\&
+\mathcal{C}\left( f|_{\mathbf{z}z_{\beta_{2}}=\mathbf{c}1},
\bar f|_{\mathbf{z}z_{\beta_{2}}=(c\oplus1)0} \right)(s)\\&
-\mathcal{C}\left( f|_{\mathbf{z}z_{\beta_{2}}=\mathbf{c}1},
\bar f|_{\mathbf{z}z_{\beta_{2}}=(c\oplus1)1} \right)(s).
    \end{aligned}
\end{equation}
So from (\ref{eqn29}), (\ref{eqn32}), (\ref{eqn30}) and (\ref{eqn33}) we get the value of $M$ in (\ref{eqn27}) as
\begin{equation}
    \begin{aligned}\label{eqn34}
      &M=2\sum_{\mathbf{c}}\left(\mathcal{C}\left( f|_{\mathbf{z}z_{\beta_{2}}=\mathbf{c}0},
\bar f|_{\mathbf{z}z_{\beta_{2}}=(c\oplus1)0} \right)(s)\right.\\& \left.
-\mathcal{C}\left( f|_{\mathbf{z}z_{\beta_{2}}=\mathbf{c}1},
\bar f|_{\mathbf{z}z_{\beta_{2}}=(c\oplus1)1} \right)(s)\right)\cdot
2^k \cdot (-1)^{\left(\mathbf{n} \oplus \mathbf{b}^{\prime}\right) \cdot \mathbf{c}}.
    \end{aligned}
\end{equation}
Since $G(Q|_\mathbf{z=c})$ is a path, so for some permutation $\pi$ of $\{0,1,\hdots,m-k-5\}$ and $c_{\alpha},c\in \mathbb{Z}_q$ the function $f|_\mathbf{z=c}$ obtained by substituting $\mathbf{z=c}$ in $f$ should be of the form
\begin{equation}\label{eqn35}
\begin{aligned}
    &f|_\mathbf{z=c}= \sum_{\alpha=0}^{m-k-6}z_{\pi(\alpha)}z_{\pi(\alpha+1)}+\sum_{\alpha=0}^{m-k-5}c_{\alpha}z_{\pi(\alpha)}+c\\&
    +\!z_{\pi(m-k-5)} \left(\bar z_{m-1}\!\left(z_{m-2}\bar z_{m-3}\bar z_{m-4}\!+\!z_{m-2}z_{m-3}\right)\!+\!z_{m-1} \right. \\& \left.\bar z_{m-2}\bar z_{m-3} \right)+ \bar z_{m-1}\left(\bar z_{m-4}\left(z_{m-3}+z_{m-2}\right)+z_{m-2}z_{m-3}\right).
    \end{aligned}
\end{equation}
Let $h{1}$ and $h{2}$ be the function obtained from $f$ by substituting $\mathbf{z}=\mathbf{c}$, $z_{\beta_{2}}=0$ and  $\mathbf{z}=\mathbf{c}$, $z_{\beta_{2}}=1$ respectively.  Further without loss of generality let $\beta_2=\pi(0)$. Then both the function can be expressed as 
\begin{equation}\label{eqn36}
    h_1=\sum_{\alpha=1}^{m-k-6}z_{\pi(\alpha)}z_{\pi(\alpha+1)}+\sum_{{\alpha=0}\atop \pi(\alpha)\neq 0}^{m-k-5}c_{\alpha}z_{\pi(\alpha)}+c+R,
\end{equation}
\begin{equation}\label{eqn37}
 \text{and}~   h_2=h_1+z_{\pi(1)}+c_0.
\end{equation}
The functions $h{1}$ and $h{2}$ give non-zero components of the complex vectors $\mathbf{e_1}=f|_{\mathbf{z} z_{\beta_{2}}=\mathbf{c} 0}$ and $\mathbf{e_2}=f|_{\mathbf{z} z_{\beta_{2}}=\mathbf{c} 1}$ respectively.
%The nonzero components of the complex vectors $\mathbf{a}=f|_{\mathbf{x} x_{\beta_{2}}=\mathbf{c} 0}$ and $\mathbf{b}=f|_{\mathbf{x} x_{\beta_{2}}=\mathbf{c} 1}$ are given by the functions $h{1}$ and $h{2}$ respectively.
Similarly, $\bar{h_2}$ and $\bar{h_1}$ give the non-zero components of the vector $\bar{f}|_{\mathbf{z} z_{\beta_{2}}=(\mathbf{c} \oplus \mathbf{1}) 0}$ and $\bar{f}|_{\mathbf{z} z_{\beta_{2}}=(\mathbf{c} \oplus \mathbf{1}) 1}$   respectively. For any complex-valued sequences $\mathbf{e_1}$ and $\mathbf{e_2}$ the following identity holds
\begin{equation}
    \mathcal{C}\left(\mathbf{e_1},\mathbf{{\bar{e}_2}}\right)(s)= \mathcal{C}\left(\mathbf{e_2},\mathbf{{\bar{e}_1}}\right)(s).
\end{equation}
Using the above identity, we get,
\begin{equation}\label{eqn38}
    \begin{aligned}
        &\mathcal{C}\left( f|_{\mathbf{z}z_{\beta_{2}}=\mathbf{c}0},
\bar f|_{\mathbf{z}z_{\beta_{2}}=(c\oplus1)0} \right)(s)\\& 
=\mathcal{C}\left( f|_{\mathbf{z}z_{\beta_{2}}=\mathbf{c}1},
\bar f|_{\mathbf{z}z_{\beta_{2}}=(c\oplus1)1} \right)(s),
    \end{aligned}
\end{equation}
which shows that $M$ in (\ref{eqn34}) is zero, and hence the result follows from (\ref{eqn27}).
\end{IEEEproof}

\begin{example}
Let us consider the set $S_t$ for $0\leq t<4$ as defined in \textit{Example} \ref{ex2}. Now for the same GBF defined in \textit{Example} \ref{ex1}, using \textit{Lemma} \ref{lem3} construct a MOCS $\bar {S_t}$ $(0\leq t <4)$ of length 160 as
\begin{equation}
        \left\{\bar f+a_0\bar z_0+a_1\bar z_3+ n_0 \bar z_0+n_1\bar z_3+\bar a z_1: a,a_0,a_1 \in \{0,1\}\right\},
\end{equation}
where $t=n_02^0+n_12^1$. Then from \textit{Theorem} \ref{thm3} 
\begin{equation}
\left\{S_t: 0\leq t<4\right\}
    \cup\left\{\bar S_t: 0\leq t<4\right\},
    \end{equation}
    is a $(8,8,160)$-CCC.   
\end{example}

%\textcolor{red}
\begin{table*}[h]
%\tiny
\centering
\caption{Comparison of the proposed CCC construction with \cite{arthina,sarkar,ccgrmc,parampalli,lying2,shing2}}
\resizebox{\textwidth}{!}{
\begin{tabular}{|l|l|l|l|l|l|}
 \hline

     Ref. & Parameters & Phase & Based on   & Length(N) & Constraints  %&\makecell{Column sequence\\ PMEPR is upper\\ bounded by }                                                                                                                
     \\ \hline
     
     \cite{parampalli}& $(2^{k+1},2^{k+1},N)$ & $q\geq2$, $q$ is even & GBF of order $>2$ & $2^m$ & $m,k\in \mathbb{Z}^+,~m>1 $ %& $2^{k+1}$  
     \\ \hline
     
         \cite{arthina} & $(2^{k+1},2^{k+1},N)$   & $q\geq2 $, $q$ is even & GBF of order $2$&  $2^m$       &$m,k\in \mathbb{Z}^+,~m>1 $ %& $2^{k+1}$                                               
         \\ \hline
       %\cite{das1} &  $(M,M,N)$   & $q\geq 2 $  & PU matrix  &   $M^m$  &   $m\geq 1$ & $M$    \\ \hline
       
       %\cite{das2} & $(M,M,N)$  &  $q\geq 2$ & PU matrix & $P^m$    &  $m\geq 1$, $P|M$      & $M$             \\ \hline
        %\cite{das3}             & Indirect & q            &              &                                              \\ \hline

     \cite{ccgrmc} & $(2^k,2^k,N)$ & $q\geq 2$, $q$ is even  & GBF of order $>2 $ &$2^m$ & $k,m\in \mathbb{Z}^+$, $m\geq1,k\leq m$
      \\ \hline
      
      \cite{lying2}& $(2^{k+1},2^{k+1},N)$ & $q\geq2$, $q$ is even & GBF of order $2$ & $2^m$ & $m,k\in \mathbb{Z}^+,m>1,1\leq k \leq m-1$ \\ \hline
 \cite{shing2} & $(2^k,2^k,N)$ & $q\geq 2$, $q$ is even & GBF of order $2$ & $2^m$ & $m,k\in \mathbb{Z}^+$, $m\geq3,1\leq k \leq m$ \\ \hline 
 
 \cite{sarkar}   &\makecell{$(M,M,N)$\\ $M=p_1p_2\hdots p_k$}  &  $q=\text{lcm}(p_1,p_2,\hdots,p_k,r)$  & MVF of order $2$     & $p_1^{m_1}p_2^{m_2}\cdots p_k^{m_k}$ & $r,m_i\in \mathbb{Z}^+$, $p_i$ is prime, $1\leq i \leq k$,        \\ \hline 
      
\multirow{2}{*}{\textit{Theorem} \ref{thm3}}   & \multirow{2}{*}{$(2^{k+1},2^{k+1},N)$} & \multirow{2}{*}{$2$} & \multirow{2}{*}{GBF of order $> 2$ }    &$2^{m-1}+2^{m-3}$        &$m,k\in \mathbb{Z}^+,m\geq 5$   \\
\cline{5-6} & & & & $2^{m-1}+2^{m-2}+2^{m-4}$ & $m,k\in \mathbb{Z}^+$, $m\geq 6$
\\ \hline
\end{tabular}}\label{table 2}
\end{table*}
In \textit{TABLE} \ref{table 2}, the proposed construction of CCC is compared with the existing construction CCC on different parameters.
\section{Construction of Sequences of length $2^{m-1}+2^{m-2}+2^{m-4}$.}\label{sec6}
In this section, we have extended our proposed construction to provide GCPs, MOCSs and binary CCCs of length $2^{m-1}+2^{m-2}+2^{m-4}$. 

Consider an integer $m \geq 6$, 
for any $c,c_i \in \mathbb{Z}_q$, we define a function
\begin{equation}\label{eqn40}
    \mathrm{f}_1(z_0,z_1,\hdots,z_{m-6})=Q+\sum_{i=0}^{m-6}c_iz_i+c,
\end{equation}
where $Q$ is the quadratic part in variables $z_0,z_1,\hdots,z_{m-6}$.
Now, we define the GBF $\mathrm{f}:A \rightarrow \mathbb{Z}_q$ as 
\begin{equation}\label{eqn41}
    \begin{aligned}
    &\mathrm{f}=\mathrm{f}_1
    +\frac{q}{2}z_{\beta_1}\\&\left(\bar z_{m-1}\bar z_{m-2}+\bar z_{m-1}z_{m-2}\left(\bar z_{m-3}\!+\!z_{m-3} \bar z_{m-4}z_{m-5}\right)\right)\\&+\frac{q}{2}\left(\bar z_{m-1}\bar z_{m-2}\left(\bar z_{m-3}z_{m-4}z_{m-5}+z_{m-3}\bar z_{m-4}\bar z_{m-5}\right)\right. \\& \left. +\!\bar z_{m-1}z_{m-2}\left(\bar z_{m-3}\bar z_{m-5}\!+\!z_{m-3}\bar z_{m-4}\!+\!z_{m-4}\left(\bar z_{m-3}z_{m-5} \right. \right. \right. \\& \left. \left. \left.+z_{m-5}\bar z_{m-5}\right)\right)+z_{m-1}\bar z_{m-2}\left(\bar z_{m-4}\bar z_{m-5}+z_{m-3}z_{m-4}\right.\right.\\& \left.\left.+\bar z_{m-3}z_{m-4}z_{m-5}+z_{m-1}z_{m-2}\bar z_{m-3}\bar z_{m-4}\right)\right),
    \end{aligned}
\end{equation}
where $A=\left\{\mathbf{0}_m,\mathbf{1}_m,\hdots(\mathbf{2^{m-1}+2^{m-2}+2^{m-4}-1})_m\right\}$. Also we define the GBF $\mathrm{\bar f} :B \rightarrow \mathbb{Z}_2$ as
\begin{equation}
\mathrm{\bar f}(z_0,z_1,\hdots,z_{m-1})=\mathrm{f}(\bar z_0,\bar z_1,\hdots,\bar z_{m-1}),
\end{equation}
 where $B=\{0,1\}^m \setminus \left\{\mathbf{0}_m,\mathbf{1}_m,\hdots,(\mathbf{2^{m-3}+2^{m-4}-1})_m \right\}$ and $\bar z_i=1-z_i.$ Now, by replacing the GBF $f$ used in the above \textit{Theorems}, by the function $\mathrm{f}$ defined in (\ref{eqn41}), %and using $L=2^{m-3}+2^{m-4}$,
 we can generate GCP, CS and CCC of length $2^{m-1}+2^{m-2}+2^{m-4}$ ($m\geq 6)$, from \textit{Theorem} \ref{thm1}, \textit{Theorem} \ref{thm2}, \textit{Theorem} \ref{thm3}, respectively.
 \begin{remark}
 The direct construction of MOCSs of length $2^{m-1}+2^{m-3}$ are available in \cite{shing} (for $t=2^{m-3}$), but MOCSs of lengths $2^{m-1}+2^{m-2}+2^{m-4}$ $(m\geq 6)$ has never been reported in the literature.
 \end{remark}
 \begin{example}
 For $m=8$ and $q=2$, let us consider the GBF
 $\mathrm{f}:\{\mathbf{0}_8,\mathbf{1}_8,\hdots,\mathbf{208}_8\} \rightarrow \mathbb{Z}_2$ defined as
\begin{equation}
    \begin{aligned}
      \mathrm{f}=&z_0z_1+z_0z_2+z_1z_2 +z_{1}\left(\bar z_{7}\bar z_{6}+\bar z_{7}z_{6}\left(\bar z_{5}+z_{5} \bar z_{4}z_{3}\right)\right)\\&+\left(\bar z_{7}\bar z_{6}\left(\bar z_{5}z_{4}z_{3}+z_{5}\bar z_{4}\bar z_{3}\right)+\bar z_{7}z_{6}\left(\bar z_{5}\bar z_{3}+z_{5}\bar z_{4}+z_{4}\left(\bar z_{5}z_{3} \right. \right. \right. \\& \left. \left. \left.+z_{3}\bar z_{3}\right)\right)+z_{7}\bar z_{6}\left(\bar z_{4}\bar z_{3}+z_{5}z_{4}+\bar z_{5}z_{4}z_{3}+z_{7}z_{6}\bar z_{5}\bar z_{4}\right)\right).  
    \end{aligned}
\end{equation}
In this example, after deleting vertex $z_2$, $\mathrm{f}$ forms a path, so the sets 
\begin{equation}
  S_0= \left\{ f+a_0 z_2+ a z_0: a,a_0, \in \{0,1\}\right\},
\end{equation}
\begin{equation}
\text{and}~    S_1= \left\{f+a_0 z_2+ a z_0+z_2: a,a_0\in \{0,1\}\right\},
\end{equation}
are MOCSs of length $208$.
Similarly the sets
\begin{equation}
  \bar{S}_0= \left\{\bar f+a_0\bar z_2+ \bar a z_0: a,a_0, \in \{0,1\}\right\},
\end{equation}
\begin{equation}
\text{and}~    \bar{S}_1= \left\{\bar f+a_0\bar z_2+ \bar a z_0+\bar z_2: a,a_0, \in \{0,1\}\right\},
\end{equation}
are MOCSs of length $208$ and hence their union i.e., the set $\{S_0,S_1,\bar{S}_0,\bar{S}_1\}$ forms a $\left(4,4,208\right)$-CCC.
 \end{example}
\section{PMEPR of MOCSs and CCCs}\label{sec7}
In this section, the row and column sequence PMEPR of the sequences generated by $\textit{Theorem}~\ref{thm2}$, $\textit{Theorem}~\ref{thm3}$ and MOCSs and CCCs constructed in the section \ref{sec6}are investigated. The PMEPR of the CCC-MC-CDMA system is determined by the column sequences of the complementary matrices when each complementary code is arranged as a matrix \cite{parampalli}. Thus, in this section, the column sequence PMEPR of constructed MOCSs and CCCs is effectively bound by 2.

Since the row sequences of $S_t$ forms a CS of size $2^k$, its PMEPR is upper bounded by $2^k$. The column sequence PMEPR of the CCC generated from $\textit{Theorem}~\ref{thm3}$ can be bounded above by $2$ by adding a suitable constant.
For GBFs $f,g$ and constants $c,c_1$ and $c_2$, it can be easily verified that $A(f+c)=A(f)$ and $C\left(f+c_1,g+c_2\right)=C(f,g)\omega^{c_1-c_2}.$\\
For a permutation $\pi'$ of $\{0,1,\hdots,k-1\}$, the set (matrix) $S_t$ of ($\ref{eqn15}$) is redefined by adding the constant $\sum_{\alpha=0}^{k-1} a_{\pi'(\alpha)}a_{\pi'(\alpha+1)}$
\begin{equation}\label{eqn39}
\begin{aligned}
    \left\{f+\sum_{\alpha=0}^{k-1} a_{\alpha} z_{p_{\alpha}}+\sum_{\alpha=0}^{k-1} n_{\alpha} z_{p_{\alpha}}+a z_{\beta_2}+\right.\\ \left.\sum_{\alpha=0}^{k-2} a_{\pi'(\alpha)}a_{\pi'(\alpha+1)}: a,a_\alpha \in \{0,1\}\right\},
    \end{aligned}
\end{equation}
where $t=\sum_{\alpha=0}^{k-1} n_{\alpha} 2^{\alpha}$.
Adding the same constant to the set $\bar S_t$ and noting that AACS remains unchanged and ACCS changes by a constant, so the new set is still a CCC with same parameters. It can be observed from ($\ref{eqn39}$) that the $i$th column of $S_t$ can be obtained by fixing $\mathbf{z}=(i_0,i_1,\hdots,i_{m-1})$, $0\leq z < 2^{m-1}+2^{m-3}$. So $i$th column of the matrix $S_t$ is dependent on a function $\phi$ defined as 
%\textcolor{red}
{
\begin{equation}
\phi(\mathbf{a})=\sum_{\alpha=0}^{k-1} a_{\pi'(\alpha)}a_{\pi'(\alpha+1)}+\sum_{\alpha=0}^{k-1} a_{\alpha} i_{p_{\alpha}}+a z_{\beta_2}+\mathcal{C},
\end{equation}
where $\mathcal{C}$ is a constant (independent of $a$).} Since any column sequence of the matrix  $S_t$ is obtained by a GBF, whose graph is a path consisting of $k$ vertices. Hence, from \cite{Davis} the $i$th column of $S_t$ is a Golay sequence, and so its PMEPR is upper bounded by 2. Similarly it can be verified that the column sequence PMEPR of $\bar S_t$ is also upper bounded by 2. So the maximum column sequence PMEPR of  $\left(2^{k+1},2^{k+1},2^{m-1}+2^{m-3}\right)$-CCC, can be suitably upper bounded by 2. The same is true for $\left(2^{k+1},2^{k+1},2^{m-1}+2^{m-2}+2^{m-4}\right)$-CCC.
\begin{remark}
There exist PU matrix based construction of CCCs of length non-power of two \cite{das1,das2,das3}, but their column sequence PMEPR are high compared to the proposed construction.
\end{remark}
\section{Conclusion}
In this paper, we have proposed a direct and generalized construction of GCP and binary CCC of non-power of two lengths by using higher-order GBFs. %and the concept of restricted Boolean functions.
 The resultant CCCs can be obtained directly from GBFs without using other tedious sequence operations. The non-power of two length binary CCCs directly constructed using GBFs finds many applications in wireless communication due to its simple modulo-2 arithmetic operation, modulation and good correlation properties. %\textcolor{red}
 {Column sequence PMEPR of the proposed CCC can be effectively reduced to be upper bounded by 2.
 The construction of MOCSs of non-power of two lengths is also provided in this paper. The proposed work  solved the open problem cited in \cite{shing,sarkar}. The work is compared with existing literature.}
\bibliographystyle{IEEEtran}
\bibliography{reference}
\end{document}